\documentclass[11pt]{article}
\usepackage{fullpage}
\usepackage{graphicx}
\usepackage{amssymb}
\usepackage{amsmath}
\usepackage{theorem}
\usepackage{natbib}
\usepackage{enumerate}

\renewcommand{\r}{\lambda}

\newcommand{\SIM}{ \hat{S}^{{ \textit{simult}}, \r}}
\newcommand{\STA}{ \hat{S}^{{ \textit{stable}}}}

\newcommand{\PIMS}{\hat{\Pi}^{{ \textit{simult}}, \r}}
\newcommand{\PIQS}{\hat{\Pi}^{{ \textit{simult}}, q}}
\newcommand{\SSM}{S_{\textit{small};\lambda}}

\newcommand{\hPi}{\hat{\Pi}}

\newcommand{\pt}{\pi_{\textit{thr}}}
\newcommand{\N}{N}

\newcommand{\rl}{\lambda}
\newcommand{\R}{\Lambda}
\newcommand{\p}{p_w}

\newcommand{\C}{C}
\newcommand{\CC}{\overline{C}}

\renewcommand{\small}{\normalsize}
\newtheorem{assum}{Assumption}
\newtheorem{remark}{Remark}
\newtheorem{lemma}{Lemma}
\newtheorem{theorem}{Theorem}

\newtheorem{definition}{Definition}
\title{Stability Selection}
\author{Nicolai Meinshausen and Peter B\"uhlmann \\ \textit{University of Oxford and ETH Z\"urich}}
\begin{document}
\maketitle
\begin{abstract} 
Estimation of structure, such as in variable selection, graphical modelling or cluster analysis is notoriously difficult, especially for high-dimensional
data. We introduce  
stability selection. 
It is based on
subsampling in combination with (high-dimensional) selection algorithms. As
such, the method is extremely general and has a very wide range of
applicability. 
Stability selection provides finite sample control for some error rates of
false discoveries and hence a transparent principle to
choose a proper amount of regularisation for structure estimation. Variable selection and structure estimation improve markedly for a range of selection methods if stability selection is applied. We prove for randomised Lasso that
stability selection will be variable selection 
consistent even if the necessary conditions needed for consistency of the
original Lasso method are violated. 
We demonstrate stability selection for variable selection and Gaussian
graphical modelling, using real and simulated data.  
\end{abstract}

\section{Introduction}

Estimation of discrete structure, such as graphs or clusters, or variable
selection is an age-old problem in statistics. It has enjoyed 
increased attention in recent years due to the massive growth of data
across many scientific disciplines. These large datasets often make
estimation of discrete structures or variable
selection imperative for improved understanding and interpretation. Most 
classical results do not cover the loosely defined
case of high-dimensional data, and it is mainly in this area where we
motivate the promising properties of our new stability selection.

In the context of regression, for
example, an active area of research is to study the $p\gg n$ case, where
the number of variables or covariates $p$ exceeds the number of
observations $n$; for an early overview see for example
\citet{geer04high}. In a similar spirit, graphical modelling with many more
nodes than sample size has been the focus of recent research, and  cluster analysis is another widely used technique to infer a discrete
structure from observed data. 

Challenges with estimation of discrete structures include computational
aspects, since corresponding optimisation problems are discrete, as well as
determining the right amount of regularisation, for example in an
asymptotic sense for consistent structure estimation. Substantial progress
has been made over the last years in developing computationally tractable
methods which have provable statistical (asymptotic) properties, even for
the high-dimensional setting with many more variables than samples. One
interesting stream of research has focused on relaxations of 
some discrete optimisation problems, for example by $\ell_1$-penalty
approaches
\citep{donoho03optimally,meinshausen04consistent,zhao05model,wainwright2006sth,yuan05model}
or greedy algorithms \citep{freund1996enb,tropp2004gga,zhang2009cfs}. The practical
usefulness of such procedures has been demonstrated in various
applications. However, the general issue of selecting a proper amount of
regularisation (for the procedures mentioned above and for many others) for
getting a right-sized structure or model has largely remained a problem
with unsatisfactory solutions.  

We address the problem of proper regularisation with a very generic
subsampling approach (bootstrapping would behave similarly). We
show that subsampling can be used to determine the amount of regularisation
such that a certain familywise type I error rate in multiple testing can be
conservatively controlled for finite sample size. Particularly for
complex, high-dimensional problems, a finite sample control is much more
valuable than an asymptotic statement with the number of observations
tending to infinity. Beyond the issue of choosing the amount of
regularisation, the subsampling approach yields a new structure estimation
or variable selection scheme.  For the more specialised case of
high-dimensional linear models, we prove what we expect in greater
generality: namely that subsampling in conjunction with $\ell_1$-penalised
estimation requires much weaker
assumptions on the design matrix for asymptotically consistent variable
selection than what is needed for the (non-subsampled) $\ell_1$-penalty
scheme. Furthermore, we show that additional improvements can be achieved
by randomising not only via subsampling but also in the selection process for
the variables, bearing some resemblance to the successful tree-based Random Forest algorithm \citep{breiman01random}. Subsampling (and bootstrapping) has been primarily used so
far for asymptotic statistical inference in terms of standard errors,
confidence intervals and statistical testing. Our work here
is of a very different nature: the marriage of subsampling and
high-dimensional selection algorithms yields finite sample familywise error
control and 
markedly improved structure estimation or selection methods. 

\subsection{Preliminaries and examples}\label{subsec.prelim}
In general, let $\beta$ be a $p$-dimensional vector, where $\beta$ is
sparse in the 
sense that $s<p$ components are non-zero. In other words, $\|\beta\|_0 = s <p$.
Denote the set of non-zero values by $S=\{k:\beta_k\neq 0\}$ and the set of
variables with vanishing coefficient by $\N=\{k:\beta_k=0\}$. The goal of
structure estimation is to infer the set $S$ from noisy
observations. 

As a first supervised example, consider data $(X^{(1)},Y^{(1)}),\ldots ,(X^{(n)},Y^{(n)})$
with univariate response variable $Y$ and $p$-dimensional covariates
$X$. We typically assume that $(X^{(i)},Y^{(i)})$'s are i.i. distributed. The
vector $\beta$ could be the coefficient vector in a linear model 
\begin{equation}\label{eq:linear} 
Y= X\beta +\varepsilon,
\end{equation}
where $Y=(Y_1,\ldots,Y_n)$, $X$ is the $n\times p$ design matrix and
$\varepsilon=(\varepsilon_1,\ldots,\varepsilon_n)$ is the random noise
whose components are independent, identically distributed. Thus, inferring
the set $S$ from data is the well-studied variable selection problem in
linear regression. A main stream of classical methods proceeds to solve
this problem by penalising the negative log-likelihood with the
$\ell_0$-norm $\|\beta\|_0$ which equals the number of non-zero components
of $\beta$. The computational task to solve such an $\ell_0$-norm penalised
optimisation problem becomes quickly unfeasible if $p$ is getting large,
even when using efficient branch and bound techniques. Alternatively, one
can relax the $\ell_0$-norm by the $\ell_1$-norm penalty. This leads to the
Lasso estimator \citep{tibshirani96regression,chen01atomic},
\begin{equation} \label{lasso}   \hat{\beta}^\rl =
  \mbox{argmin}_{\beta\in\mathbb{R}^p} \; \| Y - X\beta\|_2^2 + \rl
  \sum_{k=1}^p | \beta_k|,   
\end{equation} 
where $\rl\in \mathbb{R}^+$ is a regularisation parameter and we typically
assume that the covariates are on the same scale, i.e. $\|X_k\|_2  =
\sum_{i=1}^n (X_{k}^{(i)})^2 = 1$. 
An attractive feature of Lasso is its computational feasibility for large
$p$ since the optimisation problem in (\ref{lasso}) is convex. Furthermore,
the Lasso is able to select variables by
shrinking certain estimated coefficients exactly to 0. We can then 
estimate the set $S$ of non-zero $\beta$ coefficients by $\hat{S}^{\rl} =
\{k;\ \hat{\beta}^{\rl}_k \neq 0\}$ which involves convex optimisation
only. Substantial understanding has been gained over the last few years about
consistency of such Lasso variable selection
\citep{meinshausen04consistent,zhao05model,wainwright2006sth,yuan05model}, and we
present the details in Section \ref{subsec.randomLasso}. Among the
challenges are the  
issue of choosing a proper amount of regularisation $\rl$ for 
consistent variable selection and the fact that restrictive design conditions
are needed for asymptotically recovering the true set $S$ of relevant
covariates. 

A second example is on unsupervised Gaussian graphical modelling. The data
is assumed to be 
\begin{equation}\label{ggm}
X^{(1)},\ldots ,X^{(n)}\ \mbox{i.i.d.}\ \sim {\cal N}_d(\mu,\Sigma). 
\end{equation}
The goal is to infer conditional dependencies among the $d$ variables or
components in $X = (X_1,\ldots ,X_d)$. It is well-known that
$X_j$ and $X_k$ are conditionally dependent given all other
components $\{X_{(\ell)};\ \ell \neq j,k\}$ if and only if
$\Sigma^{-1}_{jk} \neq 0$, and we then draw an edge between nodes $j$ and
$k$ in a corresponding graph \citep{lauritzen96graphical}. The structure
estimation is 
thus on the index set ${\cal G} = \{(j,k);\ 1 \le j < k \le d\}$ which has
cardinality $p = {d \choose 2}$ (and of course, we can represent ${\cal G}$
as a $p\times 1$ vector) and the set of relevant conditional
dependencies is $S = \{(j,k) \in {\cal G};\ \Sigma^{-1}_{jk} \neq
0\}$. Similarly to the problem of variable selection in regression,
$\ell_0$-norm methods are computationally very hard and become very quickly
unfeasible for moderate or large values of $d$. A relaxation with
$\ell_1$-type penalties has also proven to be useful in this context
\citep{meinshausen04consistent}. 
A recent proposal is the graphical Lasso
\citep{friedman2007sic}: 
\begin{equation}\label{glasso}
\hat{\Theta}^{\rl} = \mbox{argmin}_{\Theta\ \mathrm{nonneg. def.}}
\{-\log(\det(\Theta)) + \mathrm{tr}(S \Theta) + \rl \sum_{j<k} |\Theta_{jk}|\}.
\end{equation}
This amounts to an $\ell_1$-penalised estimator of the Gaussian
log-likelihood, partially maximised over the mean vector $\mu$, when
minimising over all nonnegative definite symmetric matrices. The estimated
graph structure is then $\hat{S}^{\rl} = \{(j,k) \in {\cal G};\
\hat{\Theta}^{\rl}_{jk} \neq 0\}$ which involves convex optimisation only
and is computationally feasible for large values of $d$. 

Another potential area of application is clustering. Choosing the correct number of cluster is a notoriously difficult problem. Looking for clusters that are stable under perturbations or subsampling of the data can help to get a better sense of a meaningful number of clusters and to validate results.  Indeed, there has been some activity in this area, most notably in the context of \emph{consensus clustering} \citep{monti2003ccr}. For an early application see \citet{bhattacharjee2005chl}. Our proposed false discovery control can be applied to consensus clustering, yielding good estimates of the parameters of a suitable base clustering method for consensus clustering. 

\subsection{Outline}
The use of resampling for purposes of validation is certainly not new; we merely try to put it into a more formal framework and to show certain empirical and theoretical advantages of doing so. It seems difficult to give a complete coverage of all previous work in the area, as notions of stability, resampling and perturbations are very natural in the context of structure estimation and variable selection.
 We reference and compare with previous work throughout the paper.

The structure of the paper is as follows. The generic stability selection
approach, its familywise type I multiple testing error control and some
representative examples from high-dimensional linear models and Gaussian
graphical models are presented in Section~\ref{sec.stable}. A detailed asymptotic analysis of Lasso and randomised
Lasso for high-dimensional linear models is given in Section~\ref{sec.cons} and more numerical results are described in Section~\ref{sec.numeric}. After a discussion in Section~\ref{sec.disc}, we
collect all the technical proofs in the Appendix. 

\section{Stability selection}\label{sec.stable}

Stability selection is not a new variable selection technique. Its aim is
rather to enhance and improve existing methods. First, we give a general
description of stability selection and we present specific examples and
applications later.  
We assume throughout this Section \ref{sec.stable} that the data, denoted
here by $Z^{(1)},\ldots ,Z^{(n)}$, are independent and identically
distributed (e.g. $Z^{(i)} = (X^{(i)},Y^{(i)})$ with covariate $X^{(i)}$
and response $Y^{(i)}$). 

For a generic structure estimation or variable selection technique, we have a
tuning  
parameter $\r \in\R  \subseteq \mathbb{R}^+$ that determines the amount of
regularisation. This tuning parameter could be the penalty parameter in
$\ell_1$-penalised regression, see (\ref{lasso}), or in Gaussian graphical
modelling, see (\ref{glasso}); or it may be number of steps in forward
variable selection or Orthogonal Matching Pursuit \citep{mallat1993mpt} or the
number of iterations in Matching Pursuit \citep{mallat1993mpt} or Boosting
\citep{freund1996enb}; a large number of steps of iterations would
have an opposite meaning from a large penalty parameter, but this does not
cause conceptual problems. For every value $\r\in\R$, we obtain a structure
estimate $\hat{S}^\r\subseteq \{1,\ldots,p\}$. It is then of interest to
determine 
whether there exists a $\r\in\R$ such that $\hat{S}^\r$ is identical to $S$
with high probability and how to achieve that right amount of
regularisation. 

\subsection{Stability paths}
We motivate the concept of stability paths in the following, first for
regression. Stability paths are derived from the concept of regularisation
paths. A regularisation path is given by the coefficient value of each variable over all
regularisation parameters: $\{\hat{\beta}^{\rl}_k;\ \rl \in \R,\ k=1,\ldots
,p\}$. Stability paths (defined below) are, in contrast, the
\emph{probability} for each variable to be selected when randomly resampling from the data. 
For any given regularisation parameter $\r\in\R$, the selected set
$\hat{S}^\r$ is implicitly a function of the samples $I=\{1,\ldots,n\}$. We
write $\hat{S}^\r=\hat{S}^\r(I)$ where necessary to express this
dependence.  
\begin{definition}[Selection probabilities] 
Let $I$ be a random subsample of $\{1,\ldots,n\}$ of size $\lfloor n/2 \rfloor$, drawn without replacement. 
For every set $K \subseteq \{1,\ldots,p\}$, the probability of being in the selected set $\hat{S}^\r(I)$ is 
\begin{equation}\label{Pi} \hPi^\r_K \; =\;  P^*\big( K \subseteq \hat{S}^\r (I)\big).\end{equation}
\end{definition}

\begin{remark} The probability $P^*$ in (\ref{Pi}) is with respect to both the
  random subsampling (and other sources of randomness if $\hat{S}^\r$ is a randomised algorithm, see Section \ref{subsec.randomLasso}).
\end{remark}
The sample size of $\lfloor n/2 \rfloor$ is chosen as it resembles most
closely the bootstrap \citep{freedman77samp,buchlmann2002ab} while
allowing computationally efficient implementation. Subsampling has also
been advocated in a related context in \citet{valdar2009}. 

For every variable $k=1,\ldots,p$, the stability path is given by the
selection probabilities $\hPi_k^\r$, $\r \in\R$. It is a complement to the
usual path-plots that show the coefficients of all variables $k=1,\ldots,p$
as a function of the regularisation parameter. It can be seen in Figure
\ref{fig:stabpath} that this simple path plot is potentially very useful for improved variable selection for high-dimensional data.

In the remainder of the manuscript, we look at the selection probabilities of individual variables. The definition above covers also sets of variables. We could monitor the selection probability of a set of functionally related variables, say, by asking how often \emph{at least one} variable in this set is chosen or how often \emph{all} variables in the set are chosen.

\subsection{Example I: Variable selection in regression}\label{subsec.examp1}

\begin{figure}
\begin{center}
\includegraphics[width=0.95\textwidth]{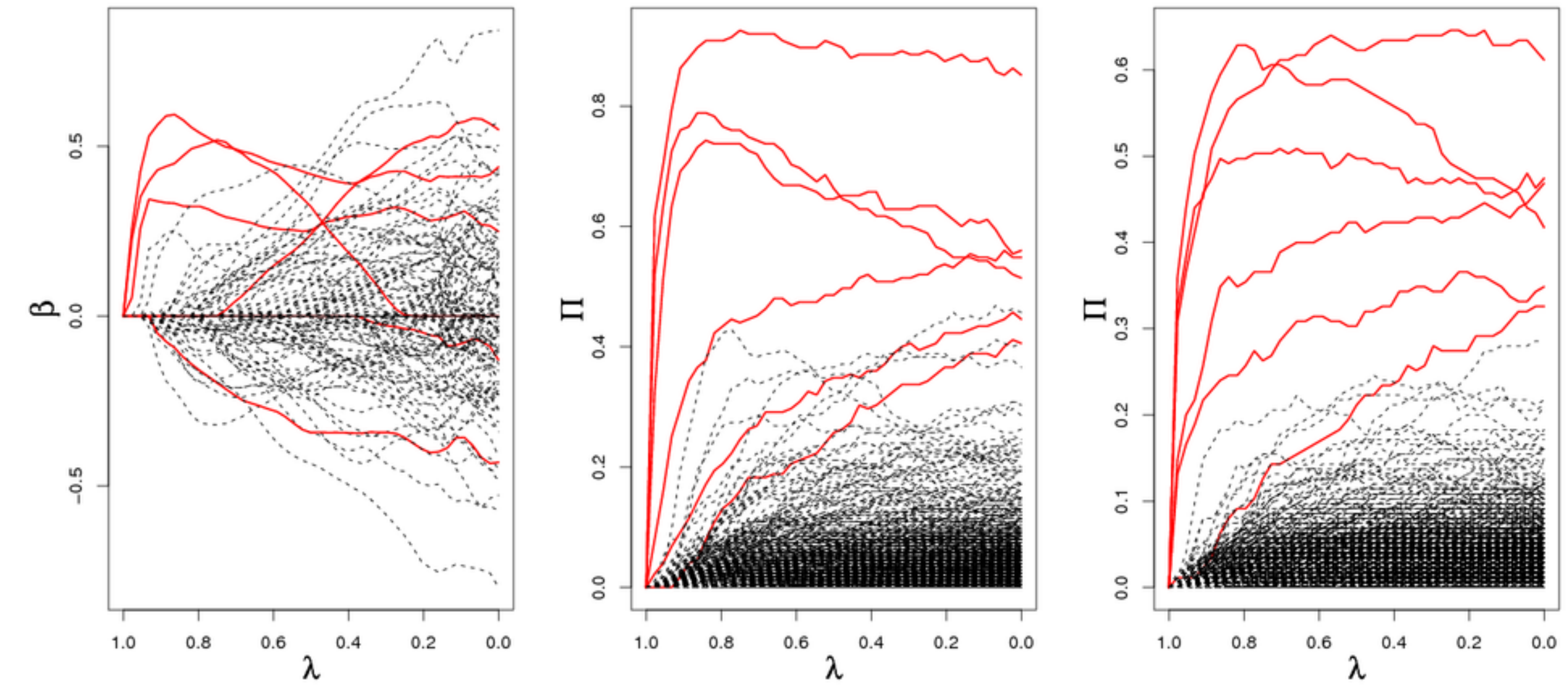}
\end{center}\vspace{-0.8cm}

{\caption{ \textit{ \small  \label{fig:stabpath}  Left: The Lasso path for
      the vitamin gene-expression dataset. The paths of the 6 non-permuted
      genes are plotted as solid, red lines, while the paths of the 4082
      permuted genes are shown as broken, black lines.  Selecting a model
      with all 6 unpermuted genes invariably means selecting a large number
      of irrelevant noise variables.  Middle: the stability path of
      Lasso. The first 4 variables chosen with stability selection are
      truly non-permuted variables. Right: The stability path for the
      `randomised Lasso' with weakness $\alpha=0.2$, introduced in Section
      \ref{subsec.randomLasso}. Now
      all 6 non-permuted variables are chosen before any noise variable
      enters the model. }}} 
\end{figure}
We apply stability selection to the Lasso defined in (\ref{lasso}). 
We work with a gene expression dataset for illustration which is kindly
provided by DSM Nutritional Products (Switzerland). For $n=115$ samples, there
is a continuous response variable measuring the logarithm of riboflavin
(vitamin B2) production rate of Bacillus Subtilis, and we have $p=4088$
continuous covariates measuring the logarithm of gene expressions from
essentially the whole genome of Bacillus Subtilis. 
Certain mutations of genes are thought to lead to higher vitamin
concentrations and the challenge is to identify those relevant genes via a
linear regression analysis. That is, we consider a linear model as in
(\ref{eq:linear}) and want to infer the set $S = \{k;\ \beta_k \neq 0\}$. 

Instability of the selected set of genes has been noted before \citep{eindor2005osg,michiels2005pco}, if either using marginal association or variable selection in a regression or classification model. 
\citet{davis2006rgs} are close in spirit to our approach by arguing for `consensus' gene signatures which assess the stability of selection, while \citet{zucknick2008ccg} propose to measure stability of so-called `molecular profiles' by the Jaccard index.

To see how Lasso and the related stability path cope with noise variables,
we randomly permute all but 6 of the 4088 gene expression across the
samples, using the same permutation to keep the dependence structure
between the permuted gene expressions intact. The set of 6 unpermuted genes has been chosen randomly among the 200 genes with the highest marginal association with the response. The Lasso path
$\{\hat{\beta}^{\rl};\ \rl \in \R\}$ is shown in
the left panel of Figure~\ref{fig:stabpath}, as a function
of the regularisation parameter $\r$ (rescaled so that $\lambda=1$ is the minimal $\lambda$-value for which the null model is selected and $\lambda=0$ amounts to the Basis Pursuit solution). Three of
the `relevant' 
(unpermuted) genes stand out, but all remaining three variables are hidden
within the paths of noise (permuted) genes. The
middle panel of Figure~\ref{fig:stabpath} shows the stability path. At least four
relevant variables stand out 
much clearer now than they did in the regularisation path plot. The
right panel shows the 
stability plot for randomised Lasso which will be introduced in Section
\ref{subsec.randomLasso}: now all 6 unpermuted variables stand
above the permuted variables and the separation between (potentially)
relevant variables and irrelevant variables is even better.  

Choosing the right regularisation parameter is very difficult for the
original path. The prediction optimal and cross-validated choice include
too many variables \citep{meinshausen04consistent,leng06nla}
and the same effect can be observed in this example, where 14 permuted
variables are included in the  model chosen by cross-validation. 
 Figure~\ref{fig:stabpath} motivates that choosing the right
regularisation parameter is much less critical for 
the stability path and that we have a better chance to select truly
relevant variables. 

\subsection{Stability selection}
In a traditional setting, variable selection would amount to choosing one
element of the set of models 
\begin{equation}\label{list}  \{\hat{S}^\r;  \;\;\r\in\R \} ,  \end{equation} 
where $\R$ is again the set of considered regularisation parameters, which
can be either continuous or discrete. There are typically two problems:
first, the correct model $S$ might not be a member of (\ref{list}). Second,
even if it is a member, it is typically very hard for high-dimensional data
to determine the right amount of regularisation $\r$ to select exactly $S$, or to
select at least a close approximation.  

With stability selection, we do not simply select one model in the list
(\ref{list}). Instead the data are perturbed (for example by subsampling)
many times and we choose all structures or variables that occur in a
large fraction of the resulting selection sets.  
\begin{definition}[Stable variables]
For a cutoff $\pt$ with $0<\pt <1$ and a set of regularisation parameters $\R$,
the set of stable variables is defined as
\begin{equation}\label{STA} \STA = \{k :\; \max_{\r\in\R} \hPi^\r_k \ge \pt \} .\end{equation}
\end{definition}

We keep variables with a high selection probability and disregard
those with low selection probabilities. The exact cutoff $\pt$ with
$0<\pt<1$ is a tuning parameter but the results vary surprisingly little
for sensible choices in a range of the cutoff. Neither do results depend
strongly on the choice of regularisation $\r$ or the regularisation region
$\R$. See Figure \ref{fig:stabpath} for an example. 

Before we present some
guidance on how to choose the cutoff parameter and the regularisation
region $\R$ below, it is worthwhile pointing out that there have been
related ideas in the literature on Bayesian model
selection. \citet{barbieri2004opm} show certain predictive optimality
results for the so-called \emph{median probability model}, consisting of
variables which have posterior probability of being in the model of 1/2 or
greater (as opposed to choosing the model with the highest posterior
probability). \citet{lee2003gsb} or \citet{sha2004bvs} are examples of more
applied papers considering Bayesian variable selection in this context. 

\subsection{Choice of regularisation and error control}
When trying to recover the set $S$, a natural goal is to include as few variables of the set $\N$ of noise variables as possible. 
The choice of the regularisation parameter is hence crucial. An advantage
of our stability selection is that the choice of the initial set of
regularisation parameters $\R$ has typically not a very strong influence on
the results, as long as $\R$ is varied with reason. Another advantage,
which we focus on below, is the ability to choose this set of regularisation
parameters in a way that guarantees, under stronger assumptions, a certain
bound on the expected number of false selections.

\begin{definition}[Additional notation]\label{def-not} Let
  $\hat{S}^\R=\cup_{\r\in\R}\hat{S}^\r$ be the set of selected structures
  or variables if varying the regularisation $\r$ in the set $\R$. Let
  $q_\R$ be the average number of selected variables, $q_\R=E(
  |\hat{S}^\R(I)| )$.  Define $V$ to be the number of
  falsely selected 
  variables with stability selection, 
\[ V = \big| \N \cap \STA \big| .\]
\end{definition}
In general, it is very hard to control $E(V)$, as the distribution of the underlying estimator $\hat{\beta}$ depends on many unknown quantities. Exact control is only possible under some simplifying assumptions. 

\begin{theorem} [Error control] \label{theo:error} Assume that the distribution of
  $\{1_{\{k\in \hat{S}^\r\}},k\in\N\}$ is exchangeable for all
  $\r\in\R$. Also, assume that the original procedure is not worse than
  random guessing, {i.e.} for any $\r \in \R$, 
\begin{equation} \label{BTRG}  \frac{ E( | S \cap \hat{S}^\r |)}{E(| \N
    \cap \hat{S}^\r |)}     \; \ge\;  \frac{| S |}{| \N|} . \end{equation}
The expected number $V$ of falsely selected variables is then bounded by 
\begin{equation}\label{EV} E(V) \quad \le\quad  \frac 1 {2\pt-1}\; \frac {
    q_\R^2} {p}.
\end{equation}
\end{theorem}
We will discuss below how to make constructive use of the value $q_\R^2$
which is in general an unknown quantity. 
The expected number of falsely selected variables is sometimes called the
per-family error rate (\textit{PFER}) or, if divided by $p$, the
per-comparison error rate (\textit{FCER}) in multiple testing
\citep{dudoit2003mht}. Choosing less variables (reducing $q_\Lambda$) or
increasing the threshold $\pt$ for selection will, unsurprisingly, reduce
the the expected number of falsely selected variables, with a minimal
achievable non-trivial value of $1/p^2$ (for $\pt=1$ and $q_\Lambda=1$) for the \textit{PFER}. This seems low enough for all
practical purposed as long as $p>10$, say. 

The involved exchangeability assumption is perhaps stronger than one would
wish, but there does not seem to be a way of getting error control in the
same generality without making similar assumptions. For regression in
(\ref{eq:linear}), the 
exchangeability assumption is fulfilled for all reasonable procedures $\hat{S}$ if the design is random and the
distribution of $\{X_k, k\in\N\}$ is exchangeable. Independence of all
variables in $\N$ is a special case. More generally, the variables could
have a joint  normal distribution with $\mathrm{Cov}(X_k,X_l) =\rho$ for
all $k,l\in \N$  
with $k\neq l$ and $ 0<\rho<1$.  For real data, we have no guarantee that
the assumption is fulfilled 
but the numerical examples in Section \ref{sec.numeric} show that the
  bound holds up very well for real data.

Note also that the assumption of exchangeability is only needed to prove
Theorem~\ref{theo:error}. All other benefits of stability selection shown
in this paper do not rely on this assumption. Besides exchangeability, we
needed another, quite harmless, assumption, namely that the original
procedure is not worse than random guessing. 
One would certainly hope that this assumption is fulfilled. If it is not,
the results below are still valid with slightly weaker constants. The
assumption seems so weak, however, that we do not pursue this further.  

The threshold value $\pt$ is a tuning parameter whose influence is very small.
For sensible values in the range of, say,  $\pt\in (0.6,0.9)$, results tend
to be very similar. Once the threshold is chosen at some default value, the
regularisation region $\R$ is determined by the desired error control.    
Specifically, for a default cutoff value $\pt=0.9$, choosing the
regularisation parameters $\R$ such that say $q_\R=\sqrt{0.8\, p}$ will
control $E(V)\le 1$; or choosing $\R$ such that
$q_\R=\sqrt{ 0.8\, \alpha\, p}$ controls the familywise error rate 
  (FWER) at level $\alpha$, {i.e.} $P(V>0)\le \alpha$.
Of course, we can proceed the other way round by fixing the regularisation
region $\R$ and choosing $\pt$ such that $E(V)$ is controlled at the
desired level.

To do this, we need knowledge about $q_\R$. This can be easily achieved by
regularisation of the selection procedure $\hat{S} = \hat{S}^q$ in terms
of the number of selected variables $q$. That is, the domain $\Lambda$ for the
regularisation parameter $\lambda$ determines the number $q$ of
selected variables, i.e. $q = q(\Lambda)$. For example, with $\ell_1$-norm
penalisation as in (\ref{lasso}) or (\ref{glasso}), the number $q$ is given
by the variables which enter first in the regularisation path when varying
from a maximal value $\lambda_{\max}$ to some minimal value
$\lambda_{min}$. Mathematically, $\lambda_{\min}$ is such that
$|\cup_{\lambda_{\max} \ge \lambda \ge \lambda_{\min}} \hat{S}^{\lambda}| \le q$. 

Without stability selection, the regularisation parameter $\r$ invariably
has to depend on the unknown noise level of the 
observations. The advantage of stability selection is that (a) exact error
control is possible, and (b) the
method works fine even though the noise level is unknown. This is a real
advantage in high-dimensional problems with $p\gg n$, as it is very hard to
estimate the noise level in these settings.

\paragraph{Pointwise Control.} For some applications, evaluation of
subsampling replicates of $\hat{S}^\r$ are already computationally very
demanding for a single value of $\r$. If this single value $\r$ is chosen
such that some overfitting occurs and the set $\hat{S}^\r$ is rather too
large, in the sense that it contains $S$ with high probability, the same
approach as above can be used and is in our experience very successful.
Results typically do not depend strongly on the utilised regularisation
$\r$. See the example below for graphical modelling. Setting $\R=\{\r\}$,
one can   
immediately transfer all results above to the case of what we call here
pointwise control. 
For methods which select structures incrementally,
{i.e.} for which 
$\hat{S}^\r \subseteq \hat{S}^{\r'}$ for all $\r\ge \r'$, pointwise control
and control with $\R=[\r,\infty)$ are equivalent since $\hat{\Pi}^\r_k$ is
then monotonically increasing with decreasing $\r$ for all
$k=1,\ldots,p$.

\subsection{Example II: Graphical modelling}
\begin{figure}
\begin{center}
\includegraphics[width=0.9\textwidth]{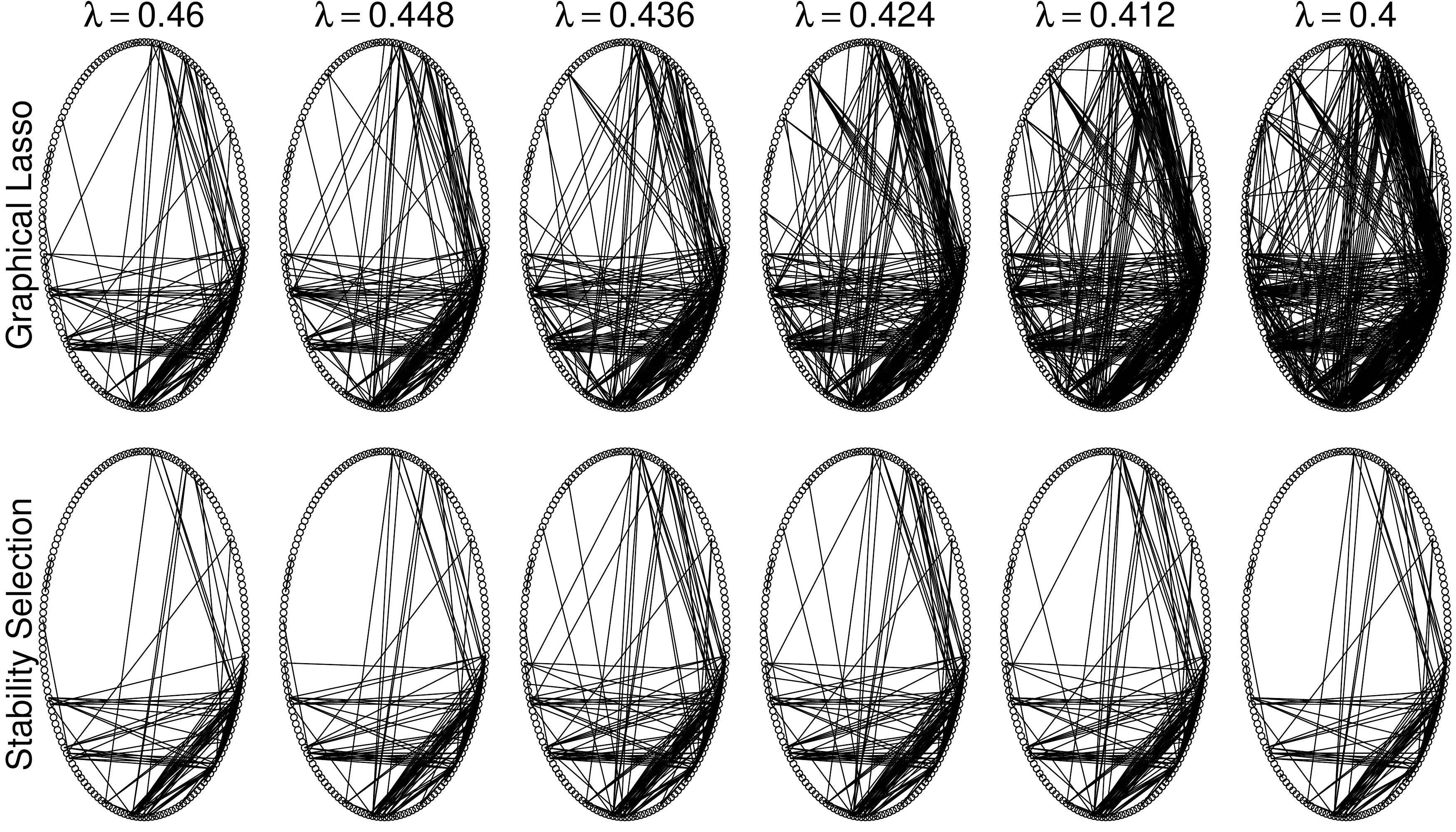}
\end{center}\vspace{-0.5cm}

{\caption{ \textit{ \small \label{fig:eggs}  Vitamin gene-expression
      dataset. The regularisation path of graphical lasso (top row) and the
      corresponding point-wise stability selected models (bottom row). } }} 
\end{figure}
\begin{figure}
\begin{center}
\includegraphics[width=0.9\textwidth]{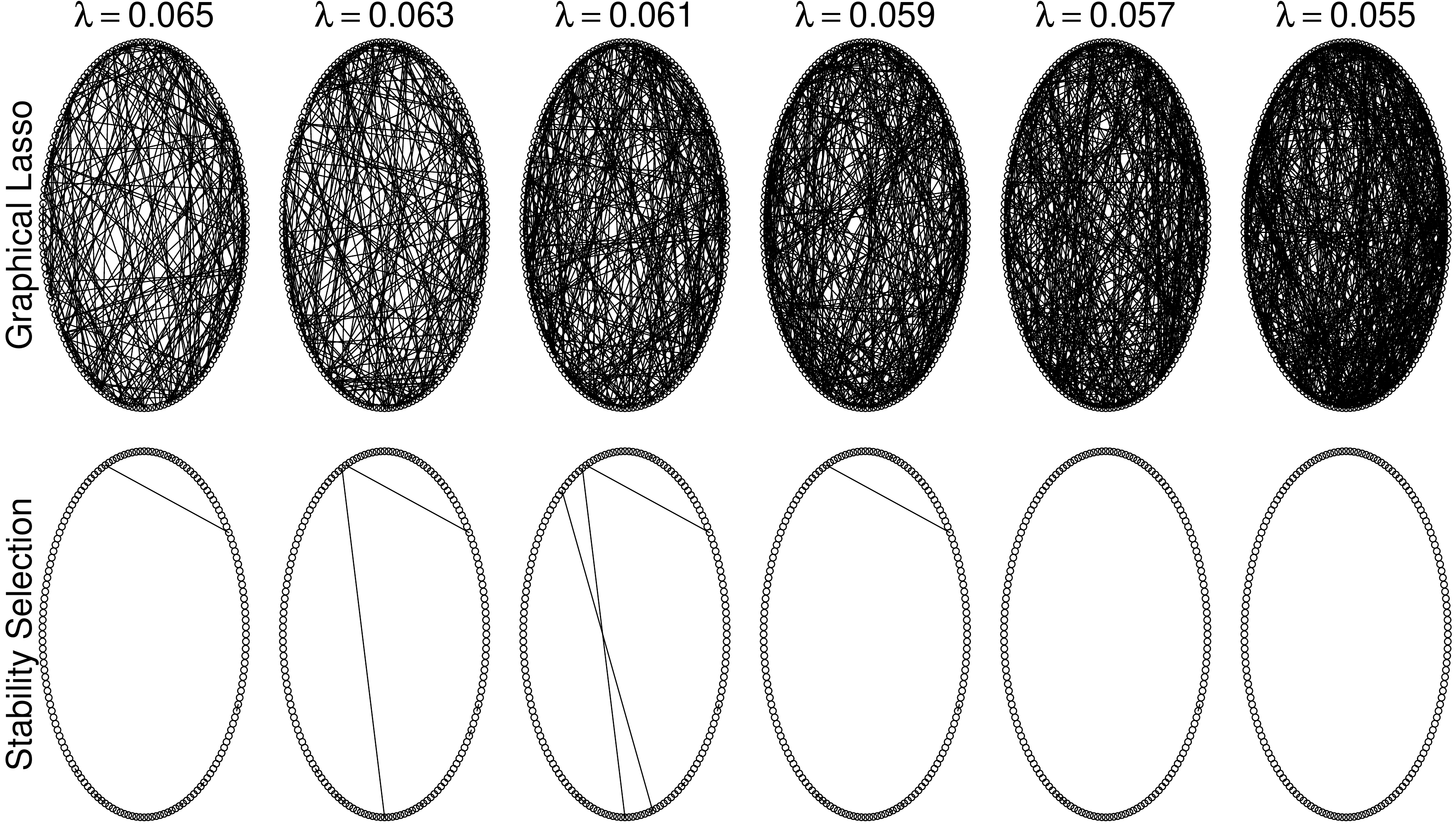}
\end{center}\vspace{-0.5cm}

{\caption{ \textit{ \small \label{fig:eggsnull}  The same plot as in
      Figure~\ref{fig:eggs} but with the variables (expression values of
      each gene) permuted independently. The empty graph is the true
      model. With stability selection, only a few errors are made, as
      guaranteed by the made error control. } }}
\end{figure}
Stability selection is also promising for graphical modelling. Here we
focus on Gaussian graphical models as described in Section
\ref{subsec.prelim} around formula (\ref{ggm}) and (\ref{glasso}). 

The pattern of non-zero
entries in the inverse covariance matrix $\Sigma^{-1}$ corresponds to the
edges between the corresponding pairs of variables in the associated graph
and is equivalent to a non-zero partial correlation (or conditional
dependence) between such pairs of variables \citep{lauritzen96graphical}.   

There has been interest recently in using $\ell_1$-penalties for model
selection in Gaussian Graphical models due to their computational
efficiency for moderate and large graphs
\citep{meinshausen04consistent,yuan05model,friedman2007sic,banerjee2008mst,bickel2008rel,rothman2008spi}. Here  
we work with the graphical Lasso \citep{friedman2007sic}, as applied
to the data from 160 randomly selected genes from the vitamin
gene-expression dataset (without the response variable) introduced in
Section \ref{subsec.examp1}. We want to infer
the set of non-zero entries in the inverse covariance matrix
$\Sigma^{-1}$. Part of the resulting regularisation path of the graphical
Lasso showing graphs for
various values of the regularisation parameter $\lambda$,
i.e. $\{\hat{S}^{\lambda};\ \lambda \in \R\}$ where $\hat{S}^{\lambda} =
\{(j,k);\ (\hat{\Sigma}^{-1})^{\rl}_{jk} \neq 0\}$, are shown in the
first row of Figure~\ref{fig:eggs}.
For reasons of display, variables
(genes) are ordered first using
hierarchical clustering and are symbolised by nodes arranged in a circle.
Stability selection is shown in the bottom row of Figure
\ref{fig:eggs}. We pursue a pointwise control approach. For each value of
$\lambda$, we select the  
threshold $\pt$ so as to guarantee $E(V)\le 30$, that is we expect fewer
than 30 wrong edges among the 12720 possible edges in the graph. The set
$\STA$ varies remarkably little for the majority of the path and the choice
of $q$ (which is implied by $\lambda$) does not seem to be critical, as
already observed for variable selection in regression.  

Next, we permute the variables (expression values) randomly, using a different
permutation for each variable (gene). The true graph is now the empty
graph. As can be 
seen from Figure~\ref{fig:eggsnull}, stability selection selects now just
very few edges or none at all (as it should). The top row shows the
corresponding graphs estimated with the graphical Lasso which yields a much
poorer selection of edges. 

\subsection{Computational requirements} 
Stability selection demands to re-run $\{\hat{S}^{\lambda};\ \lambda \in
\Lambda\}$ multiple times. Evaluating selection probabilities over 100 subsamples
seems sufficient in practice. 
The algorithmic complexity of Lasso in (\ref{lasso}) or in
(\ref{randomisedlasso}) below is of the order $O(np\min\{n,p\})$, see
\citet{efron04least}. In the $p>n$ regime, running the full Lasso
path on subsamples of size $n/2$ is hence a quarter of the cost of running
the algorithm on the full dataset and running 100 simulations is 25 times
the cost of running a single fit on the full dataset. This cost could be
compared with the cost of cross-validation, as this is what one has to
resort to often in practice to select the regularisation parameter. Running
10-fold cross-validation uses approximately $10\cdot 0.9^2=8.1$ as many
computational resources as the single fit on the full dataset. Stability
selection is thus roughly three times more expensive than 10-fold CV. This
analysis is based on the fact that the computational complexity scales like $O(n^2)$ with the number of observations (assuming $p > n$). 
If computational costs would scale linearly with sample size (e.g. for
Lasso with $p < n$), this factor would
increase to roughly 5.5. 

Stability selection with the Lasso (using 100 subsamples) for a dataset with
$p=1000$ 
and $n=100$ takes about 10 seconds on a 2.2GHz processor, using the
implementation of \citet{friedman07fastlasso}. Computational costs of this 
order would often seem worthwhile, given the potential benefits.  

\section{Consistent variable selection}\label{sec.cons}
Stability selection is a general technique, applicable to a wide range of
applications, some of which we have discussed above. Here, we want to
discuss advantages and properties of stability selection for the specific
application of variable selection in regression with high-dimensional
data which is a well-studied topic nowadays
\citep{meinshausen04consistent,zhao05model,wainwright2006sth}. We consider
a linear model as in (\ref{eq:linear}) with Gaussian noise, 
\begin{equation} \label{Ymodel}
Y= X\beta +\varepsilon,
\end{equation}
with fixed $n\times p$ design matrix $X$ and
$\varepsilon_1,\ldots,\varepsilon_n$
i.i.d.\ $\mathcal{N}(0,\sigma^2)$. The predictor variables are normalised
with 
$\|X_k\|_2=  (\sum_{i=1}^n (X_{k}^{(i)})^2)^{1/2} = 1$ for
all $k\in\{1,\ldots,p\}$. 
We allow for high-dimensional settings where $p \gg n$.

Stability selection is attractive for two
reasons. First, the choice of a proper regularisation parameter for
variable selection is
crucial and notoriously difficult, especially because the noise level is
unknown. With stability selection, results are much less sensitive to the
choice of the regularisation. Second, we will show that stability selection
makes variable selection consistent in settings where the original methods
fail.   

We give general conditions under which consistent variable selection
is achieved with stability selection. Consistent variable selection for a procedure $\hat{S}$ is
understood to be equivalent to  
\begin{equation}\label{consvar} P( \hat{S} =S) \rightarrow 1 \qquad
  n\rightarrow\infty.\end{equation} 
It is clearly of interest to know under which conditions consistent variable
selection can be achieved. In the high-dimensional context, this  places a
restriction on the growth of the number $p$ of variables and sparsity
$|S|$, typically of the form $|S|\cdot \log p =o(n)$
\citep{meinshausen04consistent,zhao05model,wainwright2006sth}. While
this assumption is often realistic, there are stronger assumptions on the
design matrix that need to be satisfied for consistent variable
selection. For Lasso, it amounts to the `neighbourhood stability' condition
\citep{meinshausen04consistent} which is equivalent to the `irrepresentable
condition' \citep{zhao05model,zou05adaptive, yuan05model}. For Orthogonal
Matching Pursuit (which is essentially forward variable selection), the
so-called `exact recovery criterion' \citep{tropp2004gga,zhang2009cfs} is sufficient
and necessary for consistent variable selection. 

Here, we show that these conditions can be circumvented more directly by
using stability selection, also giving guidance on the proper amount of
regularisation. Due to the restricted length of the paper, we will only
discuss in detail the case of Lasso whereas the analysis of Orthogonal Matching
Pursuit is just indicated.  

An interesting aspect is that stability selection with the original
procedures alone yields often very large improvements already. Moreover,
when adding  
some extra sort of randomness in the spirit of Random Forests
\citep{breiman01random} weakens 
considerably the conditions needed for consistent variables selection as
discussed next.  

\subsection{Lasso and randomised Lasso}\label{subsec.randomLasso}

The Lasso \citep{tibshirani96regression,chen01atomic} estimator is given in
(\ref{lasso}). 
For consistent variable selection using $\hat{S}^{\lambda} = \{k;\
\hat{\beta}_k^{\lambda} \neq 0\}$, it turns
out that the design needs to satisfy the so-called 
`neighbourhood stability' condition \citep{meinshausen04consistent} which is
equivalent to the `irrepresentable condition'
\citep{zhao05model,zou05adaptive, yuan05model}:
 \begin{equation} \label{IRC}\max_{k\in N} |\mbox{sign}(\beta_S)^T  (X_S^T
  X_S)^{-1} X_S^T X_k | < 1.
\end{equation}
The condition in (\ref{IRC}) is sufficient and (almost) necessary (the word
`almost' refers to the fact that a necessary relation is using `$\le$'
instead of `$<$'). 
If this condition is violated, all one can hope for is recovery of the
regression vector $\beta$ in an $\ell_2$-sense of convergence by achieving
$\|\hat{\beta}^\rl - \beta\|_2 \rightarrow_p 0$ for
$n\rightarrow\infty$. The main assumption here are bounds on the sparse
eigenvalues as discussed below.  
This
type of $\ell_2$-convergence can be used to achieve consistent variable
selection in 
a two-stage procedure by thresholding or, preferably, the adaptive Lasso
\citep{zou05adaptive,huan2008}. The disadvantage of such a two-step procedure is the
need to choose several tuning parameters without proper guidance on how
these parameters can be chosen in practice. We propose the randomised Lasso
as an alternative. Despite its simplicity, it is
consistent for variable selection even though the `irrepresentable
condition' in (\ref{IRC}) is violated.  
 
Randomised Lasso is a new generalisation of the Lasso. While the Lasso
penalises the absolute value $|\beta_k|$ of every component with a penalty
term proportional to $\rl$, the randomised Lasso changes the penalty $\rl$
to a randomly chosen value in the range $[\rl,\rl/\alpha]$. \\

\fbox{\parbox{0.8\textwidth}{
Randomised Lasso with weakness $\alpha\in(0,1]$:
\vspace{0.2cm}

Let $W_k$ be i.i.d.\ random variables in $[\alpha,1]$ for $k=1,\ldots,p$. The randomised Lasso estimator $\hat{\beta}^{\rl,W}$ for regularisation parameter $\rl\in\mathbb{R}$ is then
\begin{equation} \label{randomisedlasso} \hat{\beta}^{\rl,W} = \mbox{argmin}_{\beta\in\mathbb{R}^p} \; \| Y - X\beta\|_2^2 + \rl \sum_{k=1}^p \frac{ |  \beta_k| }{W_k}.   \end{equation}

}}\vspace{0.3cm}

A proposal for the distribution of the weights $W_k$ is described below,
just before Theorem \ref{theo:randlasso}. The word `weakness' is borrowed
from the terminology of weak greedy algorithms \citep{tem00} which are
loosely related to our randomised Lasso. 
Implementation of (\ref{randomisedlasso}) is  straightforward by
appropriate re-scaling of the predictor variables (with scale factor $W_k$
for the $k$-th variable). Using these re-scaled variables, the standard
Lasso is solved, using for example the LARS algorithm \citep{efron04least}
or fast coordinate wise approaches
\citep{meier06group,friedman07fastlasso}. 
The perturbation of the penalty weights is reminiscent of the re-weighting
in the adaptive Lasso \citep{zou05adaptive}. Here, however, the
re-weighting is not based on any previous estimate, but is simply chosen at
random! As such, it is very simple to implement. However, it seems
nonsensical at first sight since one can surely not expect any improvement
from such a random perturbation.  
If applied only with one random perturbation, randomised Lasso is not
very useful. However, applying randomised Lasso many times and looking for
variables that are chosen often will turn out to be a very powerful
procedure. 

\paragraph{Consistency for randomised Lasso with stability selection.}
For stability selection with randomised Lasso, we can do without the
irrepresentable condition (\ref{IRC}) but need only a condition on the
sparse eigenvalues of the design
\citep{candes2005dss,geer06high,meinshausen06lassotype,bickel07dantzig},
also called 
the sparse Riesz condition in \citet{zhang06model}. 
\begin{definition}[Sparse Eigenvalues]
For any $K\subseteq\{1,\ldots,p\}$, let $X_K$ be the restriction of $X$ to columns in $K$. The minimal sparse eigenvalue $\phi_{\min}$ is then defined for $k\le p$ as 
\begin{equation}\label{phimin} \phi_{\min}(k) = \inf_{a\in\mathbb{R}^{\lceil k\rceil},K\subseteq\{1,\ldots,p\}: |K|\le \lceil k\rceil } \frac{\|X_K a\|_2 }{\|a\|_2} , \end{equation}
and analogously for the maximal sparse eigenvalue $\phi_{\max}$.
\end{definition}
We have to constrain sparse eigenvalues to succeed. 
\begin{assum}[Sparse eigenvalues]\label{assum:sparse}
There exists some $\C>1$ and some $\kappa\ge 9$ such that
\begin{equation}\label{sparseassum} \frac{\phi_{\max}( \C s^2)}{\phi^{3/2}_{\min}( \C s^2)} <  \sqrt{\C}/\kappa, \qquad s=|S|. \end{equation}
\end{assum}
This assumption (\ref{sparseassum}) is related to the sparse Riesz
condition in \citet{zhang06model}. The equivalent condition there requires
the existence of some $\CC>0$ such that 
\begin{equation}\label{condzhang} \frac{\phi_{\max}((2+4 \CC)s+1)}{\phi_{\min}((2+4 \CC)s+1)} < \CC,  \end{equation}
compare with Remark 2 in \citet{zhang06model}. This assumption essentially
requires that maximal and minimal eigenvalues, for a selection of order $s$
variables, are bounded away from 0 and $\infty$ respectively. In
comparison, our assumption is significantly stronger than (\ref{condzhang}), but at the
same time typically much weaker than the standard assumption of the
`irrepresentable condition' necessary to get results comparable
to ours.  



We have not specified the exact form of perturbations we will be using for
the randomised Lasso in (\ref{randomisedlasso}). For the following, we
consider the randomised Lasso of (\ref{randomisedlasso}), where the weights
$W_k$ are sampled independently as $W_k=\alpha$ with probability
$\p\in(0,1)$ and $W_k=1$ otherwise. Other perturbations are certainly
possible and work often just as well in practice. 

\begin{theorem}\label{theo:randlasso} 
Consider the model in (\ref{Ymodel}). For randomised Lasso, let the weakness $\alpha$ be given by $\alpha^2= \nu \phi_{\min}(m)/m  $,
 for any $\nu\in ((7/\kappa)^2,1/\sqrt{2})$, and $m=\C s^2$. Let $a_n$ be a sequence with $a_n\rightarrow\infty$ for $n\rightarrow\infty$. Let $\rl_{\min}= 2 \sigma (\sqrt{2\C} s+1) \sqrt{\log (p \vee a_n)/n } $. Assume that $p>10$ and $s\ge 7$ and that the sparse eigenvalue Assumption
 \ref{assum:sparse} is satisfied. Then there exists
 some $\delta=\delta_s \in (0,1)$ such that for all $\pt \ge 1-\delta$, stability
 selection with the randomised Lasso satisfies  on a set $\Omega_A$ with $P(\Omega_A) \ge 1- 5/(p\vee a_n)$ that no noise variables are selected,
\begin{equation}\label{negative}    
 N \cap \hat{S}^{stable}_\lambda\; =\; \emptyset,
\end{equation}
where $\hat{S}^{stable}_\lambda=\{k:\hPi^\rl_k\ge \pt\}$ with  $\lambda \ge \lambda_{\min}$. On the same set $\Omega_A$,
\begin{equation} \label{positive}
 (S \setminus \SSM) \;\subseteq\; \hat{S}^{stable}_{\lambda}
\end{equation}
where $\SSM =\{k: |\beta_k| \le \, 0.3(\C s)^{3/2} \lambda \}$. This implies that all variables with sufficiently large regression coefficient are selected. 
\end{theorem}

\begin{remark}
Under the condition that the minimal non-zero regression coefficient is bounded from below by 
$ \min_{k\in S} |\beta_k| \ge (Cs)^{3/2} (0.3\lambda),$
as a consequence of Theorem~\ref{theo:randlasso},
\[ P(S=\hat{S}^{stable}_{\lambda}) \ge 1-1/(p\vee a_n),\]
i.e.\ consistent variable selection for $p\vee a_n \rightarrow\infty$
($p \to \infty$ or $n \to \infty$) in the sense of (\ref{consvar}) even if
the irrepresentable condition (\ref{IRC}) is violated. If no such lower
bound holds, the set of selected variables might miss variables with too
small regression coefficients, which are, by definition, in the set $\SSM$. 
\end{remark}

\begin{remark}\label{rem:conserv}
Theorem \ref{theo:randlasso} is valid for all $\lambda\ge
\lambda_{\min}$. This is noteworthy as it means that even if the value of
$\lambda$ is chosen too large (i.e. considerably larger than
$\lambda_{\min}$), no noise variables will be selected (formula (\ref{negative})). Only some important
variables might be missed. This effect has been seen in the empirical
examples as  stability selection is very insensitive to the choice of
$\lambda$. In contrast, a hard-thresholded solution of the Lasso with a
value of  $\lambda$ too large will lead to the inclusion of noise
variables.
Thus, stability selection with the randomised Lasso exhibits an important
property of being conservative and guarding against false positive
selections. 
\end{remark}

\begin{remark}
Theorem \ref{theo:randlasso} is derived under random perturbations of the weights. While this achieves good empirical results, it seems more advantageous in combination with with subsampling of the data. The results extend directly to this case. Let $\tilde{\Pi}^\rl_k$ be the selection probability of variable $k\in S\setminus \SSM$, while doing both random weight perturbations and subsampling $n/2$ out of $n$ observations. The probability that $\tilde{\Pi}^\rl_k$ is above the threshold $\pt\in(0,1)$ is bounded by a Markov-type inequality from below by
\[ P( \tilde{\Pi}^\rl_k \ge \pt) \ge \frac{E(\tilde{\Pi}^\rl_k)-\pt}{1-\pt} \ge  1- \frac{5}{(p\vee a_{n/2}) (1-\pt)},\]
having used that $E(\tilde{\Pi}^\rl_k)\ge 1-5/(p\vee a_{n/2})$ as a consequence of Theorem~\ref{theo:randlasso}.
If $5/(p\vee a_{n/2})$ is sufficiently small in comparison to $1-\pt$, this
elementary inequality implies that important variables in $S\setminus \SSM$
are still chosen by stability selection (subsampling and random weights
perturbation) with very high probability. A similar argument shows that
noise variables are also still not chosen with very high
probability. Empirically, combining random weight perturbations with
subsampling yields very competitive results 
and this is what we recommend to use.
\end{remark}

There is an inherent tradeoff when choosing the weakness $\alpha$. A negative consequence of a low $\alpha$ is that the design can get closer to singularity and can thus lead to unfavourable conditioning of the weighted design matrix.  On the other hand, a low value of $\alpha$  makes it less likely that irrelevant variables are selected. This is a surprising result but rests on the fact that irrelevant variables can only be chosen if the corresponding irrepresentable condition (\ref{IRC}) is violated. By randomly perturbing the weights with a low $\alpha$, this condition is bound to fail sometimes, lowering the selection probabilities for such variables. A low value of $\alpha$ will thus help stability selection to avoid selecting noise variables with a violated irrepresentable condition (\ref{IRC}).
 In practice, choosing $\alpha$ in the range of $(0.2,0.8)$ gives very useful results.

\paragraph{Relation to other work.}
In related and very interesting  work, \citet{bach08bolasso} has proposed
`Bolasso' (for bootstrapped enhanced Lasso) and shown that using a finite
number of subsamples of the original Lasso procedure and applying basically
stability selection with $\pt=1$  yields consistent variables selection
under the condition that the penalty parameter $\lambda$ vanishes faster
than typically assumed, at rate $n^{-1/2}$, and that the model dimension
$p$ is fixed. While the latter condition could possibly be technical only,
the first distinguishes it from our results. Applying stability selection
to randomised Lasso, no false variable is selected for all sufficiently
large values of $\lambda$, see Remark \ref{rem:conserv}. In other words, if
$\lambda$ is chosen `too large' with randomised Lasso, only truly relevant
variable are chosen (though a few might be missed). If $\lambda$ is chosen
too large with Bolasso, noise variables might be picked up. Figure
\ref{fig:ex3} is a good illustration. Picking the regularisation in the
left plot (without extra randomness) to select the correct model is much
harder than in the right plot, where extra randomness is added. The same
distinction can be made with two-stage procedures like adaptive Lasso
\citep{zou05adaptive} or hard-thresholding
\citep{candes2005dss,meinshausen06lassotype}, where variables are
thresholded. Picking $\lambda$ too large (and $\lambda$ is notoriously
difficult to choose), false variables will invariably enter the model. In
contrast, stability selection with randomised Lasso is not picking wrong
variables if $\lambda$ is chosen too large.   

\subsection{Example}

\begin{figure}
\begin{center}
\includegraphics[width=0.32\textwidth]{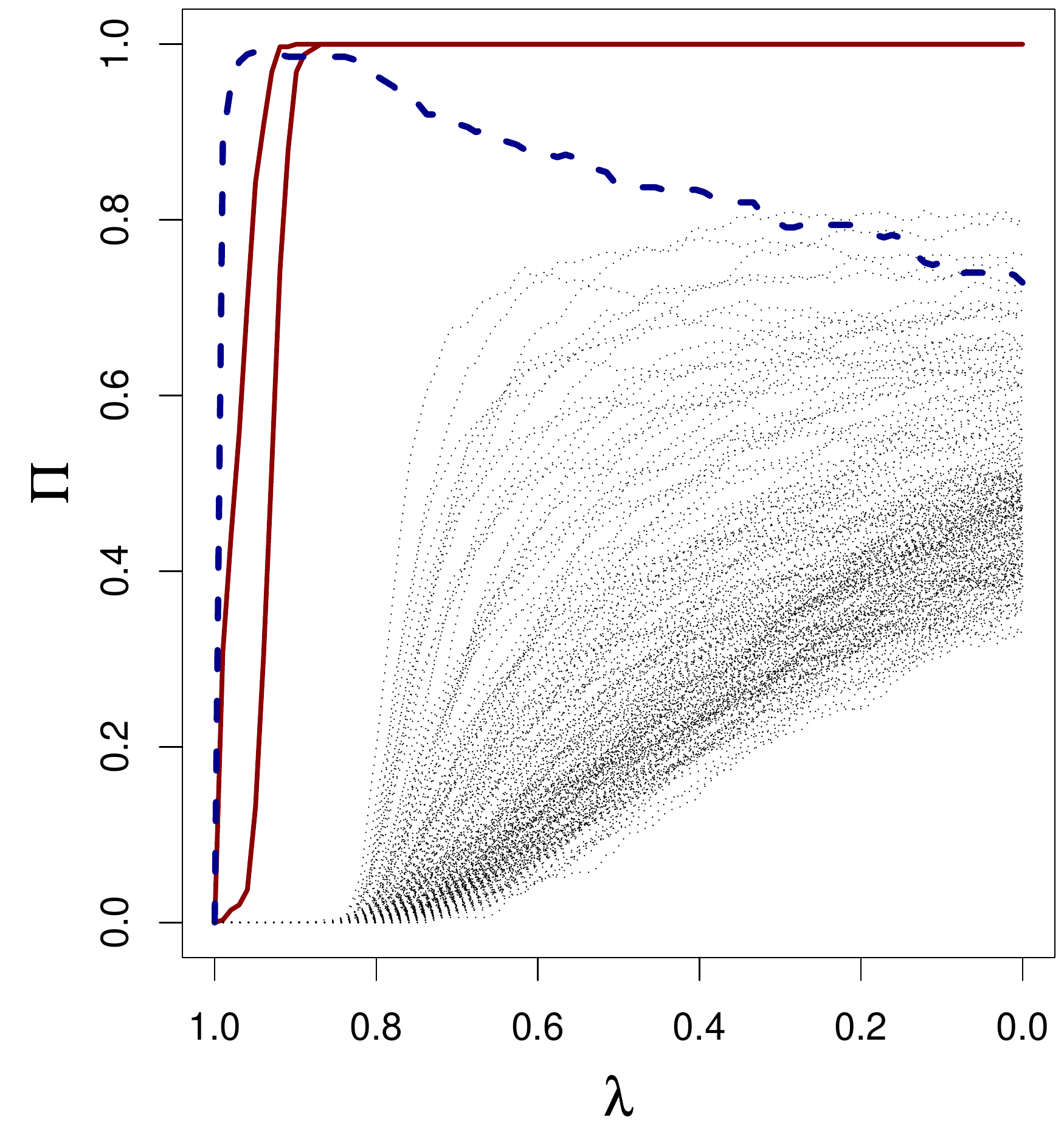}
\includegraphics[width=0.32\textwidth]{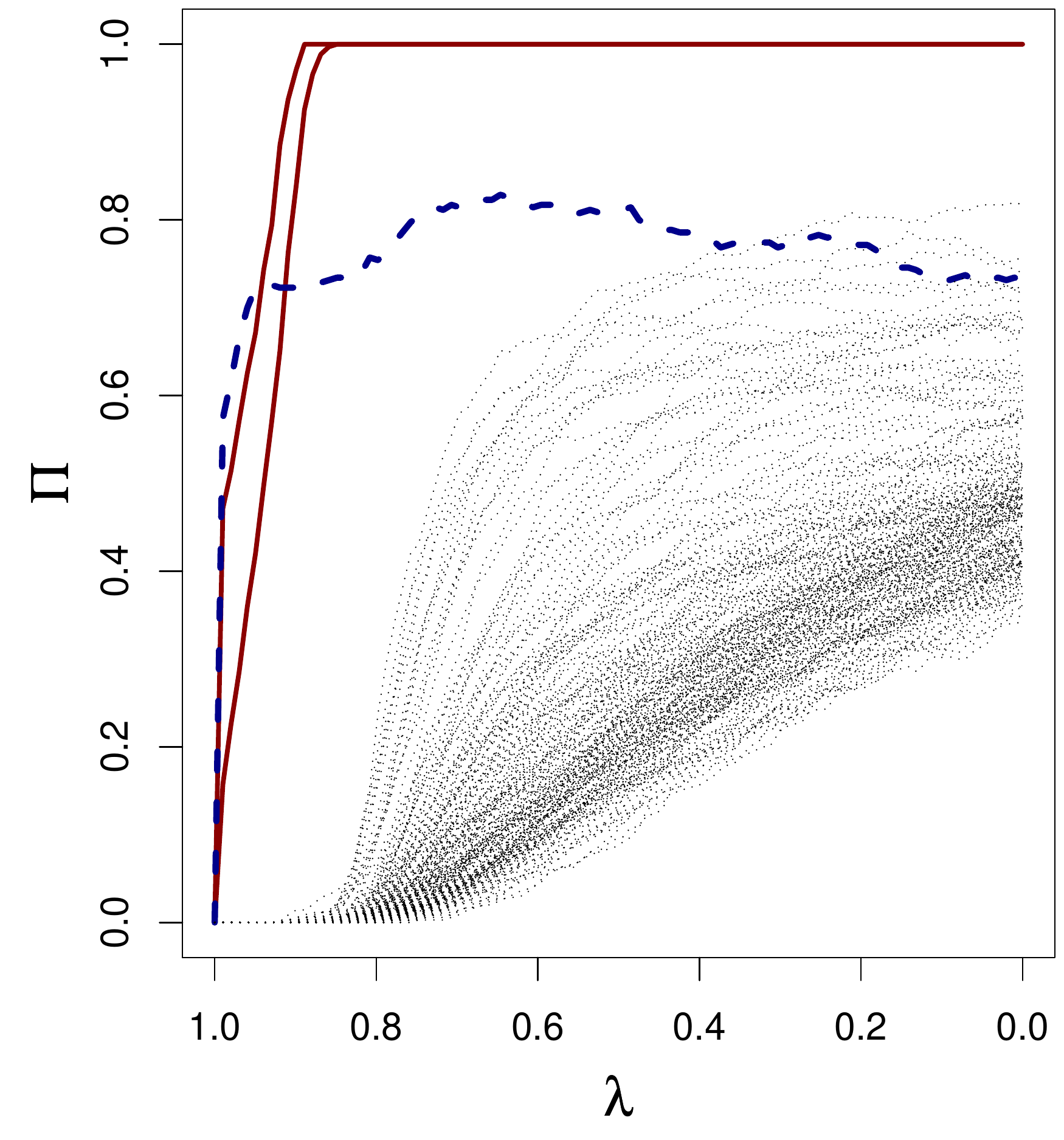}
\includegraphics[width=0.32\textwidth]{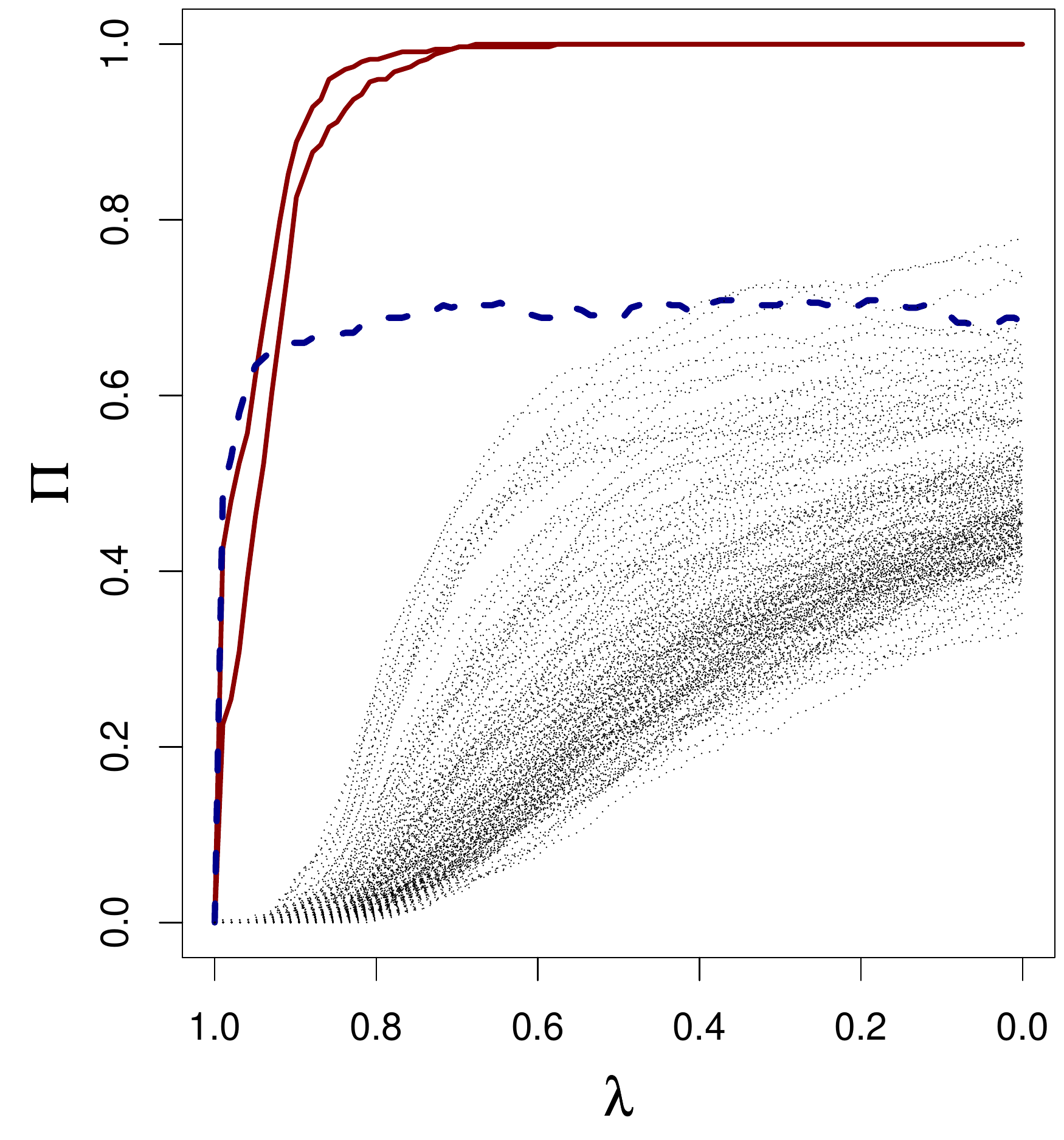}
\end{center}\vspace{-0.4cm}

{\caption{ \textit{ \small \label{fig:ex3}  The stability paths for
      randomised Lasso with stability selection using weakness parameters
      $\alpha=1$ (left panel 
      identical to the original Lasso) and $\alpha=0.5$ (middle) and
      $\alpha=0.2$ (right).  Red solid lines are the coefficients of the
      first two (relevant variables). The blue broken line is the
      coefficient of the third (irrelevant) variable and the dotted lines
      are the coefficients from all other (irrelevant)
      variables. Introducing the randomised version helps avoid choosing
      the third (irrelevant) predictor variable. }}} 
\end{figure}

We illustrate the results on randomised Lasso with a small simulation
example: $p=n=200$ and 
the predictor variables are sampled from a $\mathcal{N}(0,\Sigma)$
distribution, where $\Sigma$ is the identity matrix, except for the entries
$\Sigma_{13}=\Sigma_{23}=\rho$ and their symmetrical counterparts. We use a
regression vector $\beta=(1,1,0,0,\ldots,0)$. The response $Y$ is obtained
from the linear model $Y=X\beta+\varepsilon$ in (\ref{eq:linear}), where
$\varepsilon_1,\ldots ,\varepsilon_n$ i.i.d. 
$\mathcal{N}(0,1/4)$. For $\rho>0.5$, the
irrepresentable condition in (\ref{IRC}) 
is violated and
Lasso is not able to correctly identify the first two variables as
the truly important ones, since it always includes the third variable
superfluously 
as well. Using the randomised version for Lasso, the two
relevant variables are still chosen with probability close to 1, while the
irrelevant third variable is only chosen with much lower probability; the
corresponding probabilities are shown for randomised Lasso in Figure
\ref{fig:ex3}. This allows to separate relevant and irrelevant
variables. And indeed, the randomised Lasso is consistent under stability
selection.  

\subsection{Randomised orthogonal Matching Pursuit}
An interesting alternative to Lasso or greedy forward search in this
context are the recently proposed forward-backward search FOBA
\citep{zhang2008afb} and the MC+ algorithm \citep{Zhang07}, which both provably lead to consistent variable selection under weak conditions on sparse eigenvalues, despite being greedy solutions to non-convex optimisation problems. It will be very interesting to explore the effect of stability selection on these algorithms, but this is beyond the scope of this paper. 

Here, we look instead at orthogonal matching pursuit, a greedy forward search in the variable space. The iterative SIS procedure \citep{fan2006sis}, entails orthogonal matching pursuit as a special case. We will examine the effect of stability selection under subsampling and additional randomisation.
To have a clear definition of randomised orthogonal matching pursuit (ROMP), we define it as follows. \\

\fbox{\parbox{0.8\textwidth}{
Randomised orthogonal matching pursuit with weakness $0<\alpha<1$ and $q$ iterations.
\begin{enumerate}
\item Set $R_1=Y$. Set $m=0$ and $\hat{S}^0=\emptyset$.
\item For $m=1,\ldots,q$:
\begin{enumerate}
\item  Find $\rho_{\max}= \max_{1\le k\le p} |X_k^T R_m|$ 
\item Define $K=\{k: | X_k^T R| \ge \alpha \rho_{\max} \} $.
\item \label{en:random} Select randomly a variable $k_{sel}$ in the set $K$ and set $\hat{S}^{m} = \hat{S}^{m-1} \cup \{ k_{sel} \} $.
\item Let $R_{m+1}= Y - P_m Y$, where the projection $P_m$ is given by $X_{\hat{S}^m}(X^T_{\hat{S}^m}X_{\hat{S}^m})^{-1} X^T_{\hat{S}^m}$.
\end{enumerate}
\item Return the selected sets $\hat{S}^1\subset \hat{S}^2 \subset \ldots \subset \hat{S}^q$.
\end{enumerate}
}}\vspace{0.3cm}

A drawback of OMP is clearly that conditions for consistent variable
selection are quite strong. Following \citet{tropp2004gga}, the exact
recovery condition for OMP is defined as
\begin{equation}\label{ERC}  \max_{k\in\N} \| (X_S^T X_S)^{-1} X_S^T X_k \|_1 < 1 .\end{equation}
This is a sufficient condition for consistent variable selection. If it is
not fulfilled, there exist regression coefficients that cause OMP or its weak
variant to fail in recovery of the exact set $S$ of relevant variables.  
Surprisingly, this condition is rather similar to the irrepresentable
\citep{zhao05model} or neighbourhood stability condition
\citep{meinshausen04consistent}. 

In the spirit of Theorem \ref{theo:randlasso}, we have also a proof that
stability selection for randomised Orthogonal Matching Pursuit  (ROMP)
is asymptotically 
consistent for variable selection in linear models, even if the right hand side in (\ref{ERC}) is not bounded by 1 but instead by a possibly large constant (assuming the weakness $\alpha$ is low enough). This indicates that stability selection has a more general
potential for improved structure estimation, beyond the case for the Lasso
presented in Theorem~\ref{theo:randlasso}. It is noteworthy that our proof
involves artificial adding of noise covariates. In practice, this
seems to help often but a more involved discussion is beyond the scope of
this paper.  We will give empirical evidence for the usefulness of
stability selection under subsampling and additional randomisation for
orthogonal matching pursuit in the numerical examples below.

\section{Numerical Results}\label{sec.numeric}

\begin{figure}
\begin{center}
\includegraphics[width=0.99\textwidth]{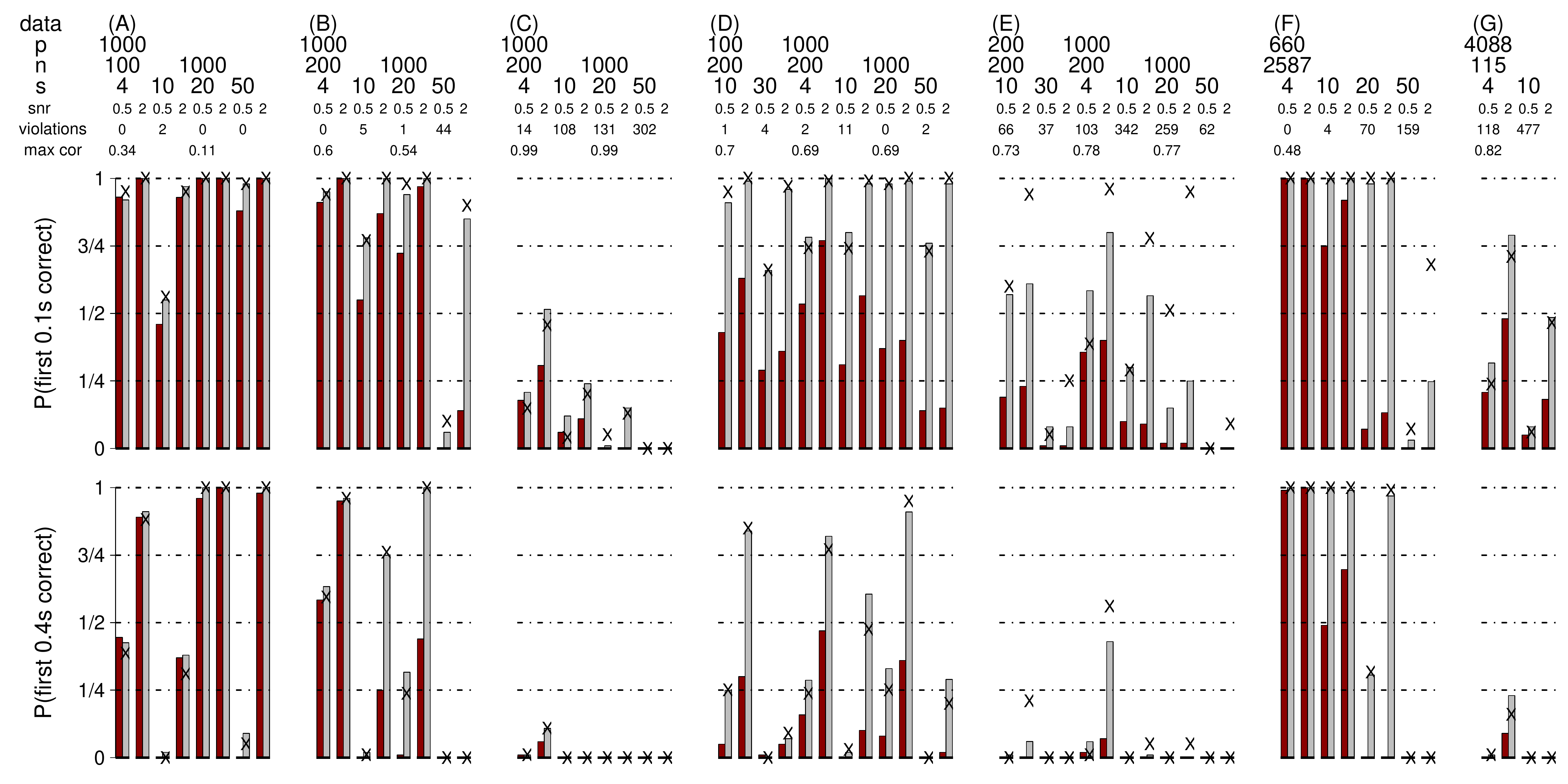}
\end{center}\vspace{-0.8cm}

{\caption{\textit{ \small \label{fig:lasso}  Probability to select $0.1s$ and $0.4s$ important variables without selecting a noise variable with the Lasso in the regression setting (dark red bar) and stability selection under subsampling (light grey bar) for the 64 different settings. Black crosses mark the result for stability selection with additional randomisation ($\alpha=0.5$). } }}
\end{figure}

\begin{figure}
\begin{center}
\includegraphics[width=0.99\textwidth]{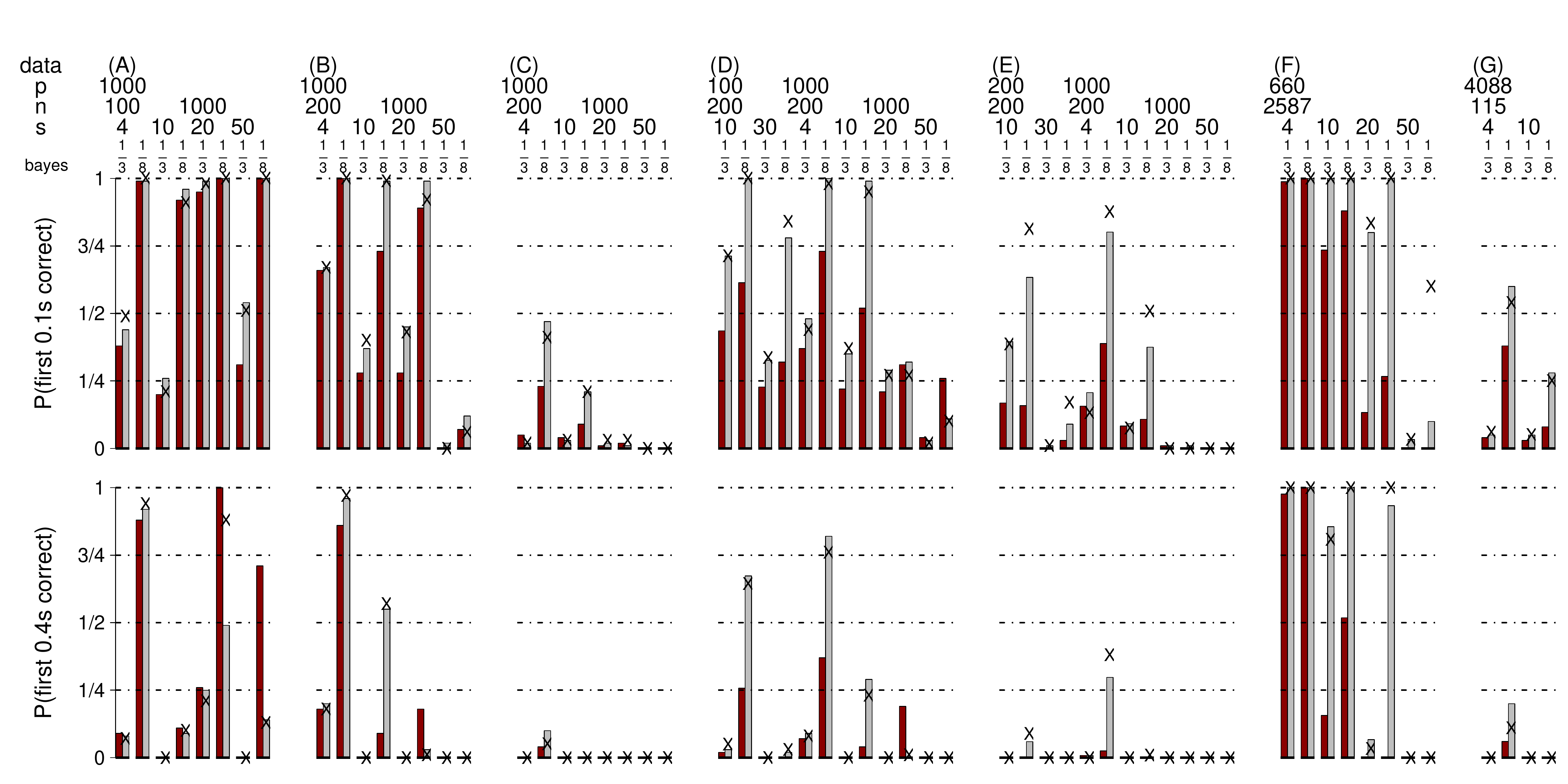}
\includegraphics[width=0.99\textwidth]{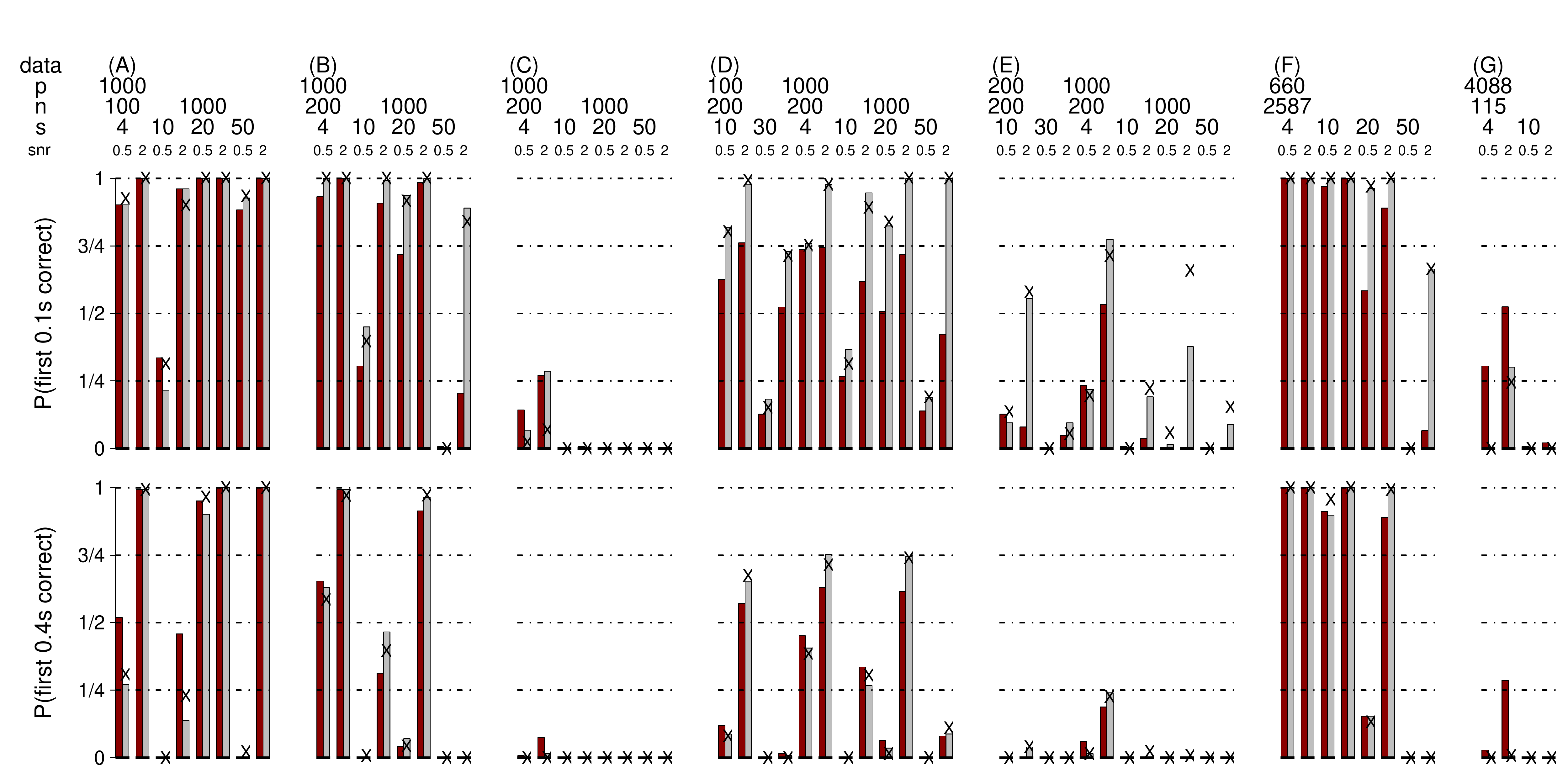}

\end{center}\vspace{-0.8cm}

{\caption{\textit{ \small \label{fig:class_omp}  The equivalent plot to Fig. \ref{lasso} for Lasso applied to classification (top two rows) and OMP applied to regression (bottom two rows). } }}
\end{figure}

\begin{figure}
\begin{center}
\includegraphics[width=0.475\textwidth]{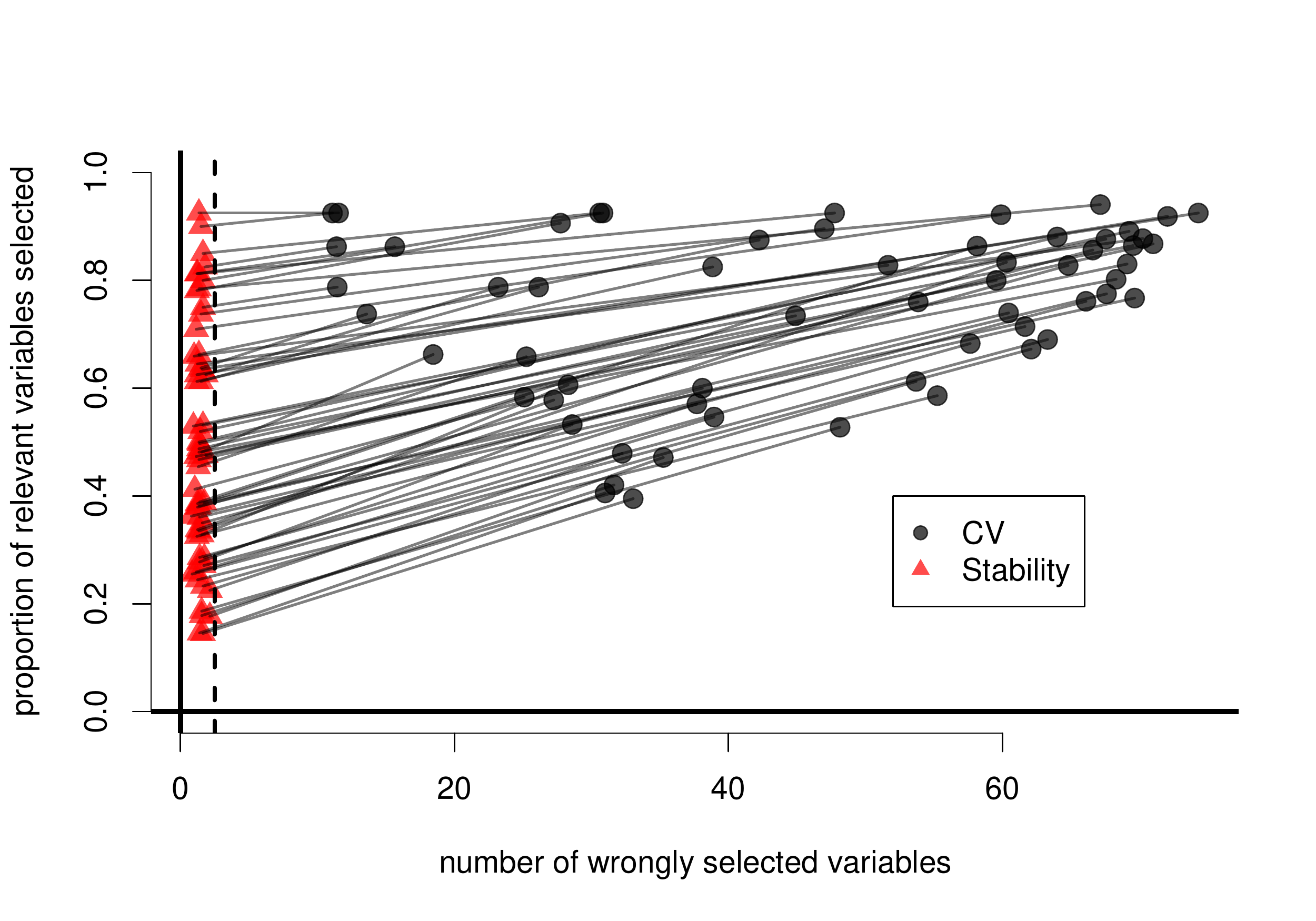}
\includegraphics[width=0.475\textwidth]{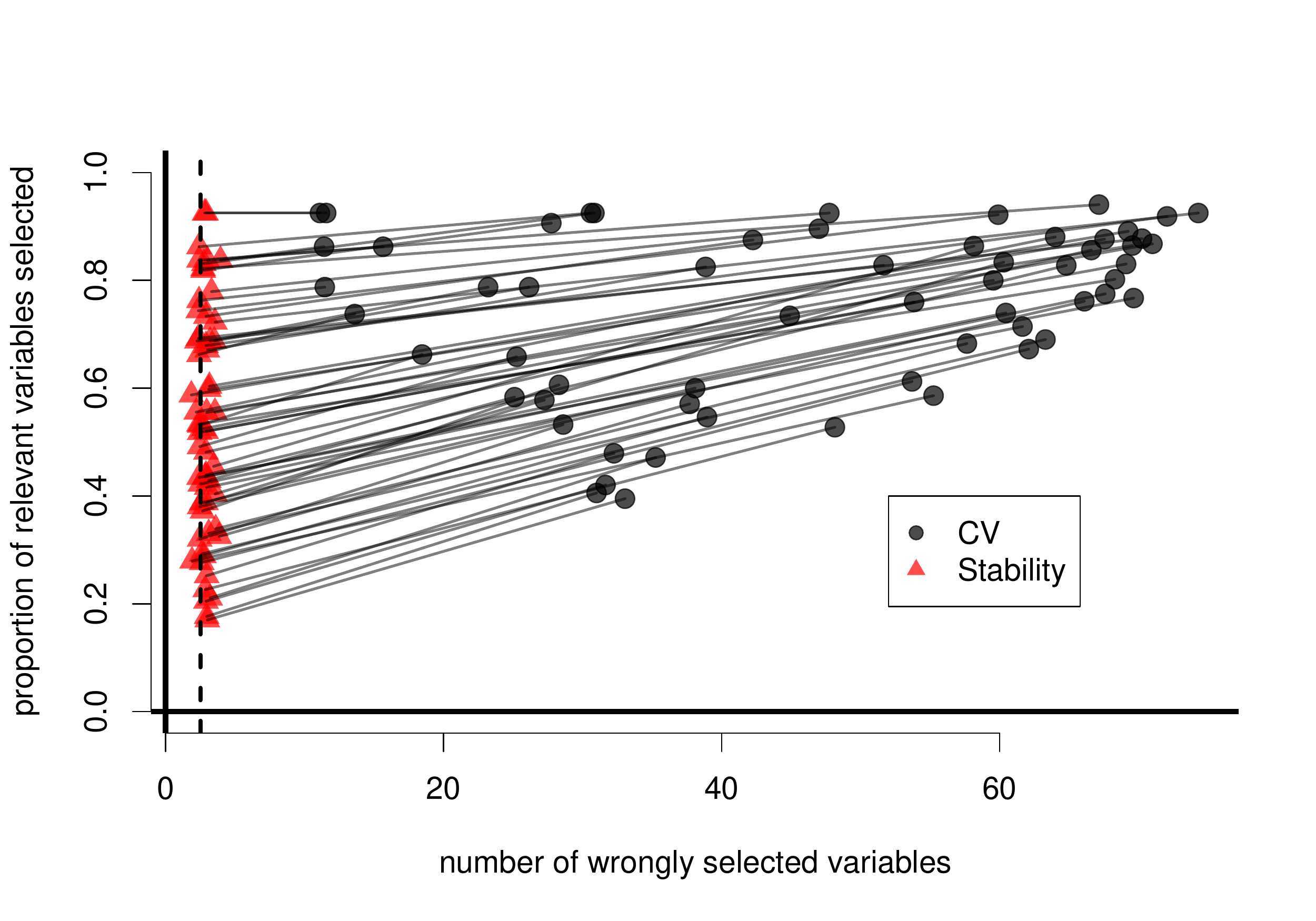}
\end{center}\vspace{-0.8cm}

{\caption{\textit{ \small \label{fig:motifs}  Comparison of stability selection with cross-validation for  the real datasets (F) and (G). The cross-validated solution (for standard Lasso) is indicated by a dot and the corresponding stability selection (for randomised Lasso, $\alpha=0.5$ on the left and $\alpha=1$ on the right) by a red triangle, showing the average proportion of correctly identified relevant variables versus the average number of falsely selected variables.  Each pair consisting of a dot and triangle corresponds to a simulation
setting (some specified SNR and $s$). The broken vertical line indicates the value at which the number of wrongly selected variables is controlled, namely $E(V)\le 2.5$. Looking at stability selection, the proportion of correctly identified relevant variables is very close to the CV-solution, while the number of falsely selected variables is reduced dramatically. } }}
\end{figure}


To investigate further the effects of stability selection, we
focus here on the application of stability selection to Lasso and 
randomised Lasso for both regression and the natural extension to
classification. The effect on OMP and randomised OMP will also be
examined. 

For regression (Lasso and OMP), we generate observations by $Y=X\beta+\varepsilon$. For classification, we use the logistic linear model under the binomial family. To generate the design matrices $X$,
we use two real and five simulated datasets,
\begin{enumerate}[(A)]
\item Independent predictor variables. All $p=1000$ predictor variables are i.i.d. standard normal distributed. Sample size $n=100$ and $n=1000$.
\item Block structure with 10 blocks. The $p=1000$-dimensional predictor variable follows a $\mathcal{N}(0,\Sigma)$ distribution, where $\Sigma_{km}=0$ for all pairs $(k,m)$ except if $\mbox{mod}_{10} k =\mbox{mod}_{10} m$, for which $\Sigma_{km}=0.5$. Sample size $n=200$ and $n=1000$.
\item Toeplitz design. The $p=1000$-dimensional predictor variable follows a $\mathcal{N}(0,\Sigma)$ distribution, where $\Sigma_{km} =\rho^{|k-m|}$ and $\rho=0.99$. Sample size $n=200$ and $n=1000$.
\item \label{D} Factor model with 2 factors. Let $\phi_1,\phi_2$ be two latent variables following i.i.d. standard normal distributions. Each predictor variable $X_k$, for $k=1,\ldots,p$, is generated as $X_k= f_{k,1}\phi_1 + f_{k,2}\phi_2 +\eta_k$, where  $f_{k,1},f_{k,1},\eta_{k}$ have i.i.d.\ standard normal distributions for all $k=1,\ldots,p$. Sample sizes are $n=200$ and $n=1000$, while $p=1000$.
\item Identical to (\ref{D}) but with 10 instead of 2 factors.
\item Motif regression dataset. A dataset ($p=660$ and $n=2587$) about finding
transcription factor binding sites (motifs) in DNA 
sequences. The real-valued predictor
variables are 
abundance scores for $p$ candidate motifs (for each of the genes). Our
dataset is from a heat-shock experiment with yeast. For a general
description and motivation about motif regression we refer to
\citet{conlon03motif}. 
\item The already mentioned vitamin gene expression data (with $p=4088$ and
$n=158$) described in Section \ref{subsec.examp1}.
\end{enumerate}
 We do not use the
response values from the real datasets, however, as we need to know which
variables are truly relevant or irrelevant. To this end, we create sparse
regression vectors by setting $\beta_k=0$ for all $k=1,\ldots,p$, except
for a randomly chosen set $S$ of  coefficients, where $\beta_k$ is chosen independently and uniformly in $[0,1]$ for all
$k\in S$. The size $s=|S|$ of the active set is varied between 4 and 50, depending on the dataset.  For regression, the noise
vector $(\varepsilon_1,\ldots,\varepsilon_n)$ is chosen i.i.d. $\mathcal{N}(0,\sigma^2/n)$, where the rescaling of the variance with $n$ is due to the
rescaling of the predictor variables to unit norm, i.e. $\|X^{(k)}\|_2 = 1$. The noise level $\sigma^2$ is chosen to achieve
signal-to-noise ratios (SNR) of $0.5$ and~$2$.  
For classification, we scale the vector $\beta$ to achieve a given Bayes misclassification rate, either $1/8$ or $1/3$. Each of the 64 scenarios is run 100 times,  once using the standard procedure (Lasso or OMP), once using stability selection with subsampling and once using stability selection with subsampling and additional randomisation ($\alpha=0.5$ for the randomised Lasso and $\alpha=0.9$ for randomised OMP). The methods are thus in total evaluated on about 20.000 simulations each.

The solution of stability selection cannot be reproduced by simply
selecting the right penalty with Lasso, since stability selection provides
a fundamentally new solution. To compare the power of both approaches, we
look at the probability that $\gamma \cdot s$ of the $s$ relevant variables can be recovered without error, where $\gamma\in\{0.1,0.4\}$. A set of $\gamma s$ variables is said to be recovered successfully for the Lasso or OMP selection, if there exists a regularisation parameter such that at least $\lceil \gamma s\rceil$ variables in $S$ have a non-zero regression coefficient and all variables in $N=\{1,\ldots,p\}\setminus S$ have a zero regression coefficient. 
For stability selection, recovery without error means that the $\lceil\gamma s\rceil$ variables with highest selection probability $\max_{\lambda\ge \lambda_{\min}} \hat{\beta}_k^\lambda$ are all in $S$. The value $\lambda_{\min}$ is chosen such that at most
$\sqrt{0.8p}$ variables are selected in the whole path of solutions for $\lambda\ge \lambda_{\min}$. Note that this notion neglects the fact that the most advantageous regularisation parameter is selected here automatically for Lasso and OMP but not for stability selection.

Results are shown in Figure \ref{fig:lasso} for Lasso applied to regression, and in Figure \ref{fig:class_omp} for Lasso applied to classification and OMP applied to regression again. In Figure \ref{fig:lasso}, we also give the median number of variables violating the irrepresentable condition (denoted by `violations') and the average of the maximal correlation between a randomly chosen variable and all other variables (`max cor') as two measures of the difficulty of the problem. 

  Stability selection identifies as many or more correct variables than the underlying method itself in all cases except for scenario (A), where it is about equivalent. That stability selection is not advantageous for scenario (A) is to be expected as the design is nearly orthogonal (very weak empirical correlations between variables), thus almost decomposing into $p$ univariate decisions and we would not expect stability selection to help in a univariate framework.

Often the gain of stability selection under subsampling is substantial, irrespective of the sparsity of the signal and the signal-to-noise-ratio. Additional randomisation helps in cases where there are many variables violating the irrepresentable condition; for example in setting (E). This is in line with our theory.

Next, we test how well the error control of Theorem
\ref{theo:error} holds up for these datasets.
For the motif regression dataset (F) and the vitamin gene expression dataset (G), Lasso is applied, with randomisation and
without. For both datasets, the signal-to-noise ratio is varied between 0.5, 1 and 2. The number of non-zero coefficients $s$ is varied in steps of 1 between 1 and 12, with a standard normal distribution for the randomly chosen non-zero coefficients. Each of the 72 settings is run 20 times.  We are interested in the comparison between the 
cross-validated solution and stability selection. For stability
selection, we chose $q_\R=\sqrt{0.8p}$ and thresholds of $\pt=0.6$,
corresponding to a control of $E(V)\le 2.5$, where $V$ is the number of
wrongly selected variables. The control is mathematically derived under the
assumption of 
exchangeability for the distribution of noise variables, see Theorem
\ref{theo:error}. This assumption is
most likely not fulfilled  for the given dataset and it is of interest to
see how well the bound holds up for real data.  
Results are shown in Figure \ref{fig:motifs}. Stability selection reduces the number of falsely selected variables dramatically, while maintaining almost the same power to detect relevant variables. The number of falsely chosen variables is remarkably well controlled at the desired level, giving empirical evidence that the derived error control is useful beyond the discussed setting of exchangeability. 
Stability selection thus helps to select a useful amount of regularisation.

\section{Discussion}\label{sec.disc}
Stability selection addresses the notoriously difficult problem of
structure estimation or variable selection, especially for high-dimensional
problems. Cross-validation fails
often for high-dimensional data, sometimes spectacularly.
Stability selection is based on subsampling in combination with
(high-dimensional) selection algorithms. The method is extremely general
and we demonstrate its applicability for variable selection in regression and
 Gaussian graphical modelling.  

Stability selection provides finite sample
familywise multiple testing error control (or control of other error rates
of false discoveries) and hence a transparent principle to
choose a proper amount of regularisation for structure estimation or variable
selection. Furthermore, the solution of stability
selection depends surprisingly little on the chosen initial regularisation. This is
an additional great benefit besides error control. 

Another property of stability selection is the improvement over
a pre-specified selection method. It is often the case that computationally
efficient algorithms for high-dimensional selection are inconsistent, even
in rather simple settings. We prove for randomised Lasso that stability selection will be variable selection
consistent even if the necessary conditions needed for consistency of the
original method are violated. And thus, stability selection will
asymptotically select the right model in scenarios where Lasso fails.

In short, stability selection is the marriage of subsampling and
high-dimensional selection algorithms, yielding finite sample familywise error
control and  markedly improved structure estimation.
Both of these main properties are demonstrated on
simulated and real data. 

\section{Appendix}\label{sec.app}

\subsection{Sample splitting}
An alternative to subsampling is sample splitting. Instead of observing if a given variable is selected for a random subsample, one can look at a random split of the data into two non-overlapping samples of equal size $\lfloor n/2 \rfloor$ and see if the variable is chosen in both sets simultaneously. 
Let $I_1$ and $I_2$ be two random subsets of $\{1,\ldots,n\}$ with $|I_i|=\lfloor n/2 \rfloor$  for $i=1,2$ and $I_1\cap I_2=\emptyset$. 
Define the simultaneously selected set as the intersection of $\hat{S}^\r(I_1)$ and $\hat{S}^\r(I_2)$,
\[ \SIM \quad =\quad  \hat{S}^\r(I_1) \,\cap\, \hat{S}^\r(I_2).\]
\begin{definition}[Simultaneous selection probability] Define the simultaneous selection probabilities $\hPi$  for any set $K\subseteq\{1,\ldots,p\}$ as
\begin{equation}\label{simult} \PIMS_K \quad =  \quad P^*( K \subseteq  \SIM ),\end{equation}
where the probability $P^*$ is with respect to the random sample splitting (and any additional randomness if $\hat{S}^\r$ is a randomised algorithm). 
\end{definition}

We work with the selection probabilities based on subsampling but the following
lemma lets us convert these probabilities easily into simultaneous
selection probabilities based on sample splitting; the latter is used for
the proof of Theorem \ref{theo:error}. The bound is rather
tight for selection probabilities close to 1.  

\begin{lemma}[Lower bound for simultaneous selection probabilities]\label{lemma:bound}
For any set $K\subseteq\{1,\ldots,p\}$, a lower bound for the simultaneous
selection probabilities is given by, for every $\omega\in\Omega$, by 
\begin{equation} \label{equiv} \PIMS_K \;\ge\;  2 \hPi^\r_K -1. 
  \end{equation}
\end{lemma}

{\small  \emph{Proof.}
Let $I_1$ and $I_2$ be the two random subsets in sample splitting of $\{1,\ldots,n\}$ with $|I_i|=\lfloor n/2 \rfloor$  for $i=1,2$ and $I_1\cap I_2=\emptyset$.  Denote by $s_K(\{1,1\})$  the probability $P^*(\{K\subseteq \hat{S}^\r(I_1)\} \cap \{K\subseteq \hat{S}^\r(I_2)\})$. Note that the two events are not independent as the probability is only with respect to a random split of the fixed samples $\{1,\ldots,n\}$ into $I_1$ and $I_2$. The probabilities $s_K(\{1,0\}),s_K(\{0,1\}),s_K(\{0,0\})$ are defined equivalently by $P^*(\{K\subseteq \hat{S}^\r(I_1)\} \cap \{K\nsubseteq \hat{S}^\r(I_2)\})$, $P^*(\{K\nsubseteq \hat{S}^\r(I_1)\} \cap \{K\subseteq \hat{S}^\r(I_2)\})$, and $P^*(\{K\nsubseteq \hat{S}^\r(I_1)\} \cap \{K\nsubseteq \hat{S}^\r(I_2)\})$. Note that $\PIMS_K = s_K(\{1,1\}) $ and 
\begin{eqnarray*}
\hPi^\r_K &=& s_K(\{1,0\}) + s_K(\{1,1\}) = s_K(\{0,1\}) + s_K(\{1,1\}) \\
1- \hPi^\r_K &=& s_K(\{0,1\}) + s_K(\{0,0\}) = s_K(\{1,0\}) + s_K(\{0,0\}) 
\end{eqnarray*} It is obvious that $s_K(\{1,0\}) = s_K(\{0,1\})$. 
As $s_K(\{0,0\})\ge 0$, it also follows that $s_K(\{1,0\}) \le 1- \hPi^\r_K$.
Hence
\[ \PIMS_K = s_K(\{1,1\}) = \hPi^\r_K - s_K(\{1,0\}) \ge 2 \hPi^\r_K -1,\]
which completes the proof.  \hfill $\Box$ }

\subsection{Proof of Theorem \ref{theo:error}}
{\small 
  The proof uses mainly Lemma \ref{lemma:markov}. We first show that $ P(k\in \hat{S}^\R) \le q_\R/p$ for all $k\in\N$, using the made definitions $\hat{S}^\R=\cup_{\r\in\R} \hat{S}^\r$ and $q_\R =E(|\hat{S}^\R|)$. Define furthermore $\N_\R = \N \cap \hat{S}^\R$ to be the set of noise variables (in $\N$) which appear in $\hat{S}^\R$ and analogously $U_\R = S\cap \hat{S}^\R$.
The expected number of falsely selected variables  can be written as $E(|\N_\R| ) =  E(|\hat{S}^\R|) - E(|U_\R|)=q_\R - E(|U_\R|)$.
Using the assumption (\ref{BTRG}) (which asserts that the method is not worse than random guessing), it follows that $E(|U_\R|)\ge E(|\N_\R|) |S|/|\N|$. Putting together, $ (1+ |S|/|\N|) E(|\N_\R|) \le q_\R $ and hence $|\N|^{-1} E(|\N_\R|) \le q_\R/p$. Using the exchangeability assumption, we have $P(k\in\hat{S}^\R) = E(|\N_\R|)/|\N|$ for all $k\in\N$ and hence, for $k\in\N$, it holds that $P(k\in \hat{S}^\R) \le q_\R/p$, as desired. Note that this result is independent of the sample size used in the construction of $\hat{S}^\r$, $\r\in\R$.  Now using Lemma \ref{lemma:markov} below, it follows that $P(\max_{\r\in\R} \PIQS_k \ge \xi) \le (q_\R/p)^2 /\xi$ for all $0<\xi<1$ and $k\in\N$. Using Lemma \ref{lemma:bound}, it follows that 
$P( \max_{\r\in\R} \hPi^\r_k \ge \pt) \le P( (\max_{\r\in\R} \PIMS+1)/2 \ge \pt) \le (q_\R/p)^2 / (2\pt -1)$. Hence $E(V) =\sum_{k\in\N} P(\max_{\r\in\R} \hPi^\r_k \ge \pt) \le q_\R^2/(p (2\pt -1))$, which completes the proof.
\hfill $\Box$ 
}

\begin{lemma}\label{lemma:markov}
Let $K\subset\{1,\ldots,p\}$ and $\hat{S}^\r$ the set of selected variables based on a sample size of $\lfloor n/2\rfloor$. If $P(K \subseteq \hat{S}^\r)\le \varepsilon$, then \[ P(\PIMS_K \ge \xi)\le \varepsilon^2/\xi .\] If $P(K \subseteq \cup_{\r\in\R} \hat{S}^\r)\le \varepsilon$ for some $\R\subseteq\mathbb{R}^+$, then \[ P(\max_{\r\in\R} \PIMS_K \ge \xi)\le \varepsilon^2/\xi .\]
\end{lemma}
{\small 
\emph{Proof.}
Let $I_1,I_2\subseteq\{1,\ldots,n\}$ be, as above, the random split of the samples $\{1,\ldots,n\}$ into two disjoint subsets, where both $|I_i|=\lfloor n/2\rfloor$ for $i=1,2$. 
Define the binary random variable $H^\r_K$ for all subsets $K\subseteq\{1,\ldots,p\}$ as
$ H^\r_K:= \mathbf{1}\big\{  K\subseteq \{\hat{S}^\r(I_1) \cap  \hat{S}^\r(I_2)  \}\big\}.$ Denote the data (the $n$ samples) by $Z$. The simultaneous selection probability $\PIMS_K$, as defined in (\ref{simult}), is then 
$ \PIMS_K = E^*(H^\r_K) = E( H^\r_K |Z),  $
where the expectation $E^*$ is with respect to the random split of the $n$ samples into sets $I_1$ and $I_2$ (and additional randomness if $\hat{S}^\r$ is a randomised algorithm).
To prove the first part, the inequality $P(K\subseteq \hat{S}^\r) \le \varepsilon$ (for a sample size $\lfloor n/2\rfloor$), implies  that
$ P(H^\r_K=1)  \le P(K\subseteq \hat{S}^\r(I_1))^2\le \varepsilon^2 $ and hence $E(H^\r_K)\le \varepsilon^2.$ Therefore, $E(H^\r_K) = E( E( H^\r_K |Z))= E(\PIMS_K)\le \varepsilon^2   $
 Using a Markov-type inequality, $ \xi P( \PIMS_K \ge \xi)   \le E(\PIMS_K) \le \varepsilon^2.$
Thus $P( \PIMS_K \ge \xi) \le \varepsilon^2/\xi $, completing the proof of the first claim. The proof of the second part follows analogously.
\hfill $\Box$}

\subsection{Proof of Theorem \ref{theo:randlasso}}

Instead of working directly with form (\ref{randomisedlasso}) of the randomised Lasso estimator, we consider the equivalent formulation of the standard Lasso estimator, where all variables have initially unit norm and are then rescaled by their random weights W.
\begin{definition}[Additional notation]\label{def:additionala}
For weights $W$ as in (\ref{randomisedlasso}), let $X^w$ be the matrix of re-scaled variables,
with $X^w_k=X_k\cdot W_k$ for each $k=1,\ldots,p$. Let $\phi_{\max}^w$ and $\phi^w_{\min}$ be the maximal and minimal eigenvalues analogous to (\ref{phimin}) for $X^w$ instead of $X$. 
\end{definition}

The proof rests mainly on the two-fold effect a weakness $\alpha<1$ has on the selection properties of the Lasso. The first effect is that the singular values of the design can be distorted if working with the reweighted variables $X^w$ instead of $X$ itself. A bound on the ratio between largest and smallest eigenvalue is derived in Lemma~\ref{lemma:trans}, effectively yielding a lower bound for useful values of $\alpha$. The following Lemma~\ref{lemma:boundedq} then asserts, for such values of $\alpha$, that the relevant variables in $S$ are chosen with high probability under any random sampling of the weights. The next Lemma~\ref{lemma:1/2} establishes the key advantage of randomised Lasso as it shows that the `irrepresentable condition' (\ref{IRC}) is sometimes fulfilled under randomly sampled weights, even though its not fulfilled for the original data. Variables which are wrongly chosen because condition (\ref{IRC}) is not satisfied for the original unweighted data will thus not be selected by stability selection.
The final result is established in Lemma~\ref{lemma:nofalse} after a bound on the noise contribution in Lemma~\ref{lemma:boundnoise}.

\begin{lemma}\label{lemma:trans}  Define $\CC$ by $(2+4\CC) s +1=C s^2$ and assume $s\ge 7$.
Let $W$ be weights generated randomly in $[\alpha,1]$, as in (\ref{randomisedlasso}), and let $X^w$ be the corresponding rescaled predictor variables, as in Definition~\ref{def:additionala}. For $\alpha^2 \ge  \nu \phi_{\min}(\C s^2)/(\C s^2)  $, with $\nu \in\mathbb{R}^+$, it holds under Assumption \ref{assum:sparse} for all random realisations $W$ that
\begin{equation}\label{boundphiw}
 \frac{\phi^w_{\max}(\C s^2)}{\phi^w_{\min}(\C s^2)}  \le  \frac{ 7 \CC}{\kappa\sqrt{\nu}}.
\end{equation}
\end{lemma}
{\small
\emph{Proof.}
Using Assumption~\ref{assum:sparse},
\[ \frac{\phi_{\max}(\C s^2)}{\phi^{3/2}_{\min}(\C s^2)}  < \frac{\sqrt{\C}}{\kappa} =  (\C s^2)^{-1/2} \frac{ ((2+4\CC)s+1)/s}{\kappa}\le (\C s^2 )^{-1/2} (3+4\CC)/\kappa , \]   
where the first inequality follows by Assumption \ref{assum:sparse}, the equality by $(2+4\CC)s+1=\C s^2$ and the second inequality by $s\ge 1$. It follows that
\begin{equation}\label{boundphi}
 \frac{\phi_{\max}(\C s^2)}{\phi_{\min}(\C s^2)}  \le \frac{3+4\CC}{\kappa} \sqrt{\frac{  \phi_{\min}(\C s^2) }{\C s^2}}.
\end{equation}
Now, let $\cal W$ be again the $p\times p$-diagonal matrix with diagonal entries ${\cal W}_{kk}=W_k$ for all $k=1,\ldots,p$ and 0 on the non-diagonal elements. Then $X^w=X\cal W$ and, taking suprema over all $\cal W$ with diagonal entries in $[\alpha,1]$,
\begin{eqnarray*} (\phi^w_{\max}(m))^2 &\le& \sup_{\cal W} \sup_{v\in\mathbb{R}^p: \|v\|_0\le m} (\| X^w v\|_2 / \|v\|_2)^2  \\ &=& \sup_{\cal W} \sup_{v\in\mathbb{R}^p:\|v\|_0\le m} ( v^T {\cal W}^TX^TX {\cal W} v ) / v^T v \le (\phi_{\max}(m))^2,
\end{eqnarray*}
where the last step follows by a change of variable transform $\tilde{v} = {\cal W} v$ and the fact that $\|v\|_0 = \| {\cal W} v\|_0$ as well as $v^Tv = \tilde{v}^T {\cal W}^{-1,T} {\cal W}^{-1} \tilde{v}$ and thus $\tilde{v}^T\tilde{v}\le   v^Tv \le \alpha^{-2} \tilde{v}^T\tilde{v}$ for all ${\cal W}$ with diagonal entries in $[\alpha,1]$. The corresponding argument for $\phi_{\min}(m)$ yields the bound $ \phi^w_{\min}(m) \ge  \alpha \phi_{\min}(m) $ for all $m\in\mathbb{N}$.
The claim (\ref{boundphiw}) follows by observing that $\CC \ge 1$ for $s\ge 7$, since $\C\ge 1$ by Assumption~\ref{assum:sparse} and hence $3+4\CC\le 7\CC$.
\hfill $\Box$
}

\begin{lemma}\label{lemma:boundedq} Let $\hat{A}^{\rl,W}$ be the set $\{k:\hat{\beta}^{\rl,W}\neq 0\}$ of selected variables of the randomised Lasso with weakness $\alpha\in (0,1]$ and randomly sampled weights $W$. Suppose that the weakness  $\alpha^2 \ge (7/\kappa)^2 \phi_{\min}(\C s^2)/(\C s^2)$.
Under the assumptions of Theorem \ref{theo:randlasso}, there exists a set $\Omega_0$  in the sample space of $Y$ with $P(Y\in\Omega_0)\ge 1-3/(p\vee a_n)$, such that for all realisations $W=w$, for $p\ge 5$, if $Y\in \Omega_0$,
\begin{equation} \label{boundedq}  |\hat{A}^{\rl,w} \cup S|\le \C s^2  \mbox{  and  }  (S\setminus \SSM) \;\subseteq\; \hat{A}^{\rl,w} ,  \end{equation}
where $\SSM$ is defined as in Theorem \ref{theo:randlasso}.
\end{lemma}
{\small 
\emph{Proof.} Follows mostly from Theorem 1 in \citet{zhang06model}. To this end, set  $c_0=0$ in their notation.  We also have $\C s^2 \le (2+4\CC)s+1$, as, by definition, $(2+4\CC)s+1=\C s^2$, as in Lemma \ref{lemma:trans}.  The quantity $C=c^*/c_*$ in \citet{zhang06model} is identical to our notation $\phi^w_{\max}(\C s^2)/\phi^w_{\min}(\C s^2)$. It is bounded for all random realisations of $W=w$, as long as $\alpha^2 \ge (7/\kappa)^2 \phi_{\min}(\C s^2)/(\C s^2)$, using Lemma \ref{lemma:trans}, by 
\[  \frac{\phi^w_{\max}( (2+4\CC)s+1)}{\phi^w_{\min}( (2+4\CC)s+1)}  \le  \CC.
 \] Hence all assumptions of Theorem 1 in \citet{zhang06model} are fulfilled, with $\eta_1=0$, for any random realisation $W=w$. Using (2.20)-(2.24) in \citet{zhang06model}, it follows that there exists a set $\Omega_0$ in the sample space of $Y$ with $P(Y\in\Omega_0) \ge 2-\exp(2/(p\vee a_n))-2/(p\vee a_n)^2\ge 1-3/(p\vee a_n)$ for all $p\ge 5$, such that if $Y\in\Omega_0$, from (2.21) in \citet{zhang06model}, 
\begin{equation}\label{use}  | \hat{A}^{\rl,w} \cup S| \le (2+4\CC)s \le \C s^2, \end{equation} and, from (2.23) in \citet{zhang06model}, 
\begin{equation}\label{use2}  \sum_{k\in S} |\beta_k|^2 1\{ k\notin \hat{A}^{\rl,w} \} \le (\frac{2}{3} \CC + \frac{28}{9} \CC^2 + \frac{16}{9} \CC^3) s \lambda^2 \le 5.6 \CC^3 s^3 \lambda^2 \le (0.3\, (\C s)^{3/2} \lambda)^2   ,\end{equation}
having used for the first inequality that, in the notation of \citet{zhang06model}, $1/(c^*c_*)\le c^*/c_*$. The $n^{-2}$ factor was omitted to account for our different normalisation. For the second inequality, we used $4\CC\le \C s$.  The last inequality implies, by definition of $\SSM$ in Theorem \ref{theo:randlasso}, that  $S\setminus \SSM \subseteq \hat{A}^{\rl,w}$, which completes the proof. 
\hfill $\Box$

}

\begin{lemma}\label{lemma:1/2} Set $m=\C s^2$. Let $k\in\{1,\ldots,p\}$ and let $K(w)\subseteq\{1,\ldots,p\}$ be a set which can depend on the random weight vector $W$. Suppose that $K(w)$ satisfies $|K(w)|\le m$ and $k\notin K(w)$ for all realisations $W=w$. Suppose furthermore that $K(w)=A$ for some $A\subseteq\{1,\ldots,p\}$ implies that $K(v) = A$ for all pairs $w,v\in\mathbb{R}^p$ of weights that fulfill $v_{j} \le w_{j}$ for all $j \in\{1,\ldots,p\}$, with equality for all $j\in A$.
Then, for $\alpha^2 \le  \phi_{min}(m)/(\sqrt{2}m) $, 
\begin{equation}
 \;P_w\big( \| ((X^w_{K(w)})^T X^w_{K(w)})^{-1} (X^w_{K(w)})^T X^w_k\|_1    \le  2^{-1/4}\big) \ge  \p(1-\p)^{m}.
\end{equation}
where the probability $P_w$ is with respect to random sampling of the weights $W$ and $p_w$ is, as above, the probability of choosing weight $\alpha$ for each variable and $1-p_w$ the probability of choosing weight~1.
\end{lemma}
{\small 
\emph{Proof.}
Let $\tilde{w}$ be the realisation of $W$ for which $\tilde{w}_k=\alpha$ and $\tilde{w}_j=1$ for all other $j\in \{1,\ldots,p\}\setminus k$. The probability of $W=\tilde{w}$ is clearly $\p(1-\p)^{p-1}$ under the used sampling scheme for the weights. Let $A:=K(\tilde{w})$ be the selected set of variables under these weights.  Let now $\mathcal{W} \subseteq \{1,\alpha\}^p$ be the set of all weights for which $w_k=\alpha$ and $w_j=1$ for all $j\in A$, and arbitrary values in $\{\alpha,1\}$ for all $w_j$ with $j\notin A\cup k$. The probability for a random weight being in this set is $P_w(w\in\mathcal{W})=\p(1-\p)^{|A|}$. By the assumption on $K$, it holds that $K(w)=A$ for all $w\in\mathcal{W}$, since $w_{j}\le \tilde{w}_{j}$ for all $j\in\{1,\ldots,p\}$ with equality for $j\in A$. For all weights $w\in\mathcal{W}$, it follows moreover that
\[  ((X^w_{A})^T X^w_{A})^{-1} (X^w_{A})^T X^w_k = \alpha (X_{A}^T X_{A})^{-1} X_{A}^T X_k .\]
Using the bound on $\alpha$, it hence only remains to be shown that, if $\|X_l\|_2=1$ for all $l\in\{1,\ldots,p\}$,  \begin{equation} \label{toshowww}  \sup_{A: |A| \le m} \; \sup_{k\notin A} \;   \| (X_{A}^TX_{A})^{-1} X_{A}^T X_k   \|^2_1 \le  m / \phi_{\min}(m).  \end{equation}
Since $\|\gamma\|_1\le \sqrt{|A|} \|\gamma\|_2$ for any vector $\gamma\in\mathbb{R}^{|A|}$, it is sufficient to show, for $\gamma:= (X_{A}^TX_{A})^{-1} X_{A}^T X_k$,
\[ \sup_{A: |A| \le m} \; \sup_{k\notin A} \;   \|  \gamma  \|^2_2 \le  1 / \phi_{\min}(m)  . \]
As $X_{A}\gamma$ is the projection of $X_k$ into the space spanned by $X_{A}$ and $\|X_k\|^2_2=1$, it holds that $\|X_{A} \gamma\|^2_2\le 1$.  Using $\|X_{S}\gamma\|_2^2 = \gamma^T (X_{A}^TX_{A}) \gamma \ge \phi_{\min} (|A|) \|\gamma\|_2^2$, it follows that $\|\gamma\|_2^2 \le 1/\phi_{\min}(|A|)$, which shows  (\ref{toshowww}) and thus completes the proof.
\hfill $\Box$ }

\begin{lemma}\label{lemma:boundnoise} Let $P_A= X_A (X_A^TX_A)^{-1} X_A^T $ be the projection into the space spanned by all variables in subset $A\subseteq\{1,\ldots,p\}$. Suppose $p>10$. Then there exists a set $\Omega_1$ with $P(\Omega_1)\ge 1-2/(p\vee a_n)$, such that for all $\omega\in\Omega_1$,
\begin{equation}
 \sup_{A:|A|\le m} \sup_{k\notin A}  | X_k^T (1-P_A) \varepsilon | < 2 \sigma(\sqrt{2m} +1) \sqrt{\log(p\vee a_n)/n}  .
\end{equation}
\end{lemma}
{\small
\emph{Proof.} 
 Let $\Omega'_1$ be the event that 
$ \max_{k\in\{1,\ldots,p\}} | X_k^T \varepsilon | \le 2\sigma \sqrt{\log(p\vee a_n)/n}.$
As entries in $\varepsilon$ are {i.i.} $\mathcal{N}(0,\sigma^2)$ distributed, $P(\Omega'_1)\ge 1- 1/(p\vee a_n)$ for all $\delta\in (0,1)$. 
Note that, for all $A\subset \{1,\ldots,p\}$ and $k\notin A$, 
$ | X_k^T P_A \varepsilon|  \le \| P_A \varepsilon\|_2 $. Define $\Omega''_1$ as \begin{equation}\label{toshowbound} \sup_{|A| \le m} \| P_A \varepsilon\|_2\le 2 \sigma\sqrt{2m \log (p\vee a_n)/n}. \end{equation}  It is now sufficient to show that $P(\Omega''_1)\ge 1-1/(p\vee a_n)$.
Showing this bound is related to a bound in \citet{zhang06model} and we repeat a similar argument. Each term $\sqrt{n} \|P_A\varepsilon\|_2/\sigma$ has a $\chi_{|A|}^2$ distribution as long as $X_A$ is of full rank $|A|$.  Hence, using the same standard tail bound as in the proof of Theorem 3 of \citet{zhang06model},
\[ P\Big( n\|P_A\varepsilon\|_2^2/\sigma^2 \ge |A| (1+4\log (p \vee a_n) )\Big) \le ( (p \vee a_n)^{-4} (1+4\log (p \vee a_n)))^{|A|/2}\le (p \vee a_n)^{-3|A|/2},\] having used  $1+4\log (p\vee a_n)\le (p\vee a_n)$ for all $p> 10$ in the last step and thus, using ${p \choose{ |A|} }\le p^{|A|}/{|A|!}$,
\[  P(\Omega''_1)\ge 1- \sum_{|A|=2}^m {{p} \choose{ |A|}} (p\vee a_n) ^{-3|A|/2} 
\ge 1 - \sum_{|A|=2}^m (p\vee a_n)^{-|A|/2} /(|A|)! \ge 1-1/(p\vee a_n),\]
which completes the proof by setting $\Omega_1=\Omega'_1\cap \Omega''_1$ and concluding that $P(\Omega_1)\ge 1-2/(p\vee a_n)$ for all $p> 10$.
\hfill $\Box$
}

\begin{lemma}\label{lemma:nofalse}Let $\delta_w= \p(1-\p)^{\C s^2}$ and $\hPi_k^\lambda=P_w(k\in\hat{A}^{\lambda,W})$ be again the probability for variable $k$ of being in the selected subset, with respect to random sampling of the weights $W$.
Then, under the assumptions of Theorem \ref{theo:randlasso}, for all $k\notin S$ and $p>10$, there exists a set $\Omega_A$ with $P(\Omega_A) \ge 1-5/(p\vee a_n)$ such that for all $\omega\in\Omega_A$ and $\rl\ge \rl_{\min}$,
\begin{eqnarray}  \max_{k\in N}  \hPi_k^\lambda &<& 1-\delta_w \label{toshowif1} \\ \min_{k \in S\setminus \SSM} \hPi_k^\lambda &\ge & 1-\delta_w , \label{toshowif2} \end{eqnarray} where  $\SSM$ is defined as in Theorem \ref{theo:randlasso}.
\end{lemma}
{\small
\emph{Proof.} We let $\Omega_A=\Omega_0\cap \Omega_1$, where  $\Omega_0$ is the event defined in Lemma \ref{lemma:boundedq} and event $\Omega_1$ is defined in  Lemma \ref{lemma:boundnoise}. Since, using these two lemmas, \[P(\Omega_0\cap \Omega_1)\ge 1- P(\Omega_0^c) - P(\Omega_1^c) \ge 1-3/(p\vee a_n)-2/(p\vee a_n)=1-5/(p\vee a_n),\] it is sufficient to show (\ref{toshowif1}) and (\ref{toshowif2}) for all $\omega \in \Omega_0\cap \Omega_1$.
We begin with  (\ref{toshowif1}).
A variable $k\notin S$ is in the selected set $\hat{A}^{\rl,W}$ only if 
\begin{equation}\label{tmp1}  | (X^w_k)^T (Y - X^w_{-k}\hat{\beta}^{\rl,W,-k})  | \ge \lambda   ,  \end{equation}
where $\hat{\beta}^{\rl,W,-k}$ is the solution to (\ref{randomisedlasso}) with the constraint that $\hat{\beta}_k^{\rl,W,-k}=0$, comparable to the analysis in \citet{meinshausen04consistent}. Let $\hat{A}^{\rl,W,-k}:=\{j :\hat{\beta}^{\rl,W,-k}_{j}\neq 0 \}$ be the set of non-zero coefficients and  $\hat{B}^{\rl,W,-k} := \hat{A}^{\rl,W,-k}\cup S$ be the set of regression coefficients which are either truly non-zero or estimated as non-zero (or both). We will use $\hat{B}$ as a short-hand notation for $\hat{B}^{\rl,W,-k}$. Let $P^w_{\hat{B}}$ be the projection operator into the space spanned by all variables in the set $\hat{B}$. For all $W=w$, this is identical to
\[ P^w_{\hat{B}} = X^w_{\hat{B}} ((X^w_{\hat{B}})^TX^w_{\hat{B}})^{-1} X^w_{\hat{B}} = X_{\hat{B}} (X_{\hat{B}}^TX_{\hat{B}})^{-1} X_{\hat{B}} =P_{\hat{B}}.\]
Then, splitting the term $(X_k^w)^T(Y - X^w_{-k}\hat{\beta}^{\rl,W,-k})$ in (\ref{tmp1}) into the two terms 
\begin{equation}\label{twoterms} (X_k^w)^T(1-P^w_{\hat{B}})  (Y - X^w_{-k}\hat{\beta}^{\rl,W,-k}) \; +\; (X_k^w)^TP^w_{\hat{B}} (Y - X^w_{-k}\hat{\beta}^{\rl,W,-k}),\end{equation}
it holds for the right term in (\ref{twoterms}) that 
\begin{eqnarray*} (X^w_k)^T P^w_{\hat{B}} (Y - X^w_{-k}\hat{\beta}^{\rl,W,-k}) &\le& (X^w_k)^T X^w_{\hat{B}} ((X^w_{\hat{B}})^TX^w_{\hat{B}})^{-1} \mbox{sign}({\hat{\beta}^{\rl,W,-k}}) \lambda \\ &\le& \| ((X^w_{\hat{B}})^TX^w_{\hat{B}})^{-1} (X^w_{\hat{B}}) ^T X_k^w\|_1 \lambda.\end{eqnarray*}
Looking at the left term in (\ref{twoterms}), since $Y\in\Omega_0$, we know by Lemma \ref{lemma:boundedq} that $|\hat{B}| \le \C s^2$ and, by definition of $\hat{B}$ above, $S\subseteq \hat{B}$. Thus  the left term in (\ref{twoterms}) is bounded from above by
\begin{eqnarray*}  (X_k^w)^T(1-P^w_{\hat{B}}) \varepsilon &\le&  \sup_{A:|A|\le \C s^2} \; \sup_{k\notin A}\; | (X_k)^T (1- P_{\hat{B}}) \varepsilon| \cdot \|X_k^w\|_2 / \|X_k\|_2 \\ &<&  \lambda_{\min}   \|X_k^w\|_2 / \|X_k\|_2,\end{eqnarray*}
having used Lemma \ref{lemma:boundnoise} in the last step and $\lambda_{\min}=2 \sigma(\sqrt{2\C}s +1) \sqrt{\log(p\vee a_n)/n}$.
Putting together, the two terms in (\ref{twoterms}) are bounded, for all $\omega\in \Omega_0\cap \Omega_1$, by
\[  \lambda_{\min}  \|X_k^w\|_2 / \|X_k\|_2 + \| ((X^w_{\hat{B}})^TX^w_{\hat{B}})^{-1} (X^w_{\hat{B}}) ^T X_k^w\|_1 \lambda.   \] We now apply  Lemma \ref{lemma:1/2} to the rightmost term. The set $\hat{B}$ is a function of the weight vector and satisfies for every realisation of the observations $Y\in\Omega_0$ the conditions in Lemma \ref{lemma:1/2} on the set $K(w)$. First,  $|\hat{B}|\le \C s^2$. Second, by definition of $\hat{B}$ above,  $k\notin \hat{B}$ for all weights $w$. Third, it follows by the KKT conditions for Lasso that the set of non-zero coefficients of $\hat{\beta}^{\rl,w,-k}$ and $\hat{\beta}^{\rl,v,-k}$ is identical for two weight vectors $w$ and $v$, as long $v_{j}=w_{j}$ for all $j\in \hat{A}^{\rl,W,-k}$ and $v_{j}\le w_{j}$ for all $j \notin \hat{A}^{\rl,W,-k}$ (increasing the penalty on zero coefficients will leave them at zero, if the penalty for non-zero coefficients is kept constant). 
Hence there exists 
 a set $\Omega_w$  in the sample space of $W$ with $P_w(\Omega_w)\ge 1-\delta_w$ such that $\| ((X^w_{\hat{B}})^TX^w_{\hat{B}})^{-1} (X^w_{\hat{B}}) ^T X_k^w\|_1\le 2^{-1/4}$.  Moreover, for the same set $\Omega_w$, we have $ \|X_k^w\|_2 / \|X_k\|_2=\alpha \le 1/s \le 1/7$. Hence, for all $\omega\in \Omega_0\cap \Omega_1$ and, for all $\omega \in \Omega_w$,  the lhs of (\ref{tmp1}) is bounded from above by $\lambda_{\min}/7 + \lambda 2^{-1/4} < \lambda$ and variable $k\notin S$ is hence not part of the set $\hat{A}^{\lambda,W}$. 
It follows that $\max_{\lambda\in\R} \hPi_k^\lambda < 1-\delta_w $ with $\delta_w=p_w(1-p_w)^{\C s^2}$ for all $k\notin S$. This completes the first part (\ref{toshowif1}) of the proof.

For the second part (\ref{toshowif2}), we need to show that, for all $\omega \in \Omega_0 \cap \Omega_1$, all variables $k$ in $S$ are chosen with probability at least $1-\delta_w$ (with respect to random sampling of the weights $W$), except possibly for variables in  $ \SSM\subseteq S$, defined in Theorem~\ref{theo:randlasso}. For all $\omega\in\Omega_0$, however, it follows directly from Lemma~\ref{lemma:boundedq} that $(S\setminus \SSM)\;  \subseteq \hat{A}^{\rl,W}$. Hence, for all $k\in S\setminus \SSM$, the selection probability satisfies
$ \hPi_k^\lambda  =1$ for all $Y\in\Omega_0$, which completes the proof.
\hfill $\Box$

} 

Since the statement in Lemma~\ref{lemma:nofalse} is a reformulation of the assertion of Theorem~\ref{theo:randlasso}, the proof of the latter is complete.

\section*{Acknowledgments} Both authors would like to thank anonymous referees for many helpful comments and suggestions which greatly helped to improve the manuscript. N.M. would like to thank FIM (Forschungsinstitut f\"ur Mathematik) at ETH Z\"urich for support and hospitality.

\end{document}